\documentclass[12pt,a4paper]{article}

\usepackage{epsfig}
\usepackage{xspace}
\usepackage{amsmath}
\usepackage{cite} 
\newlength{\abstwidth}
\newlength{\captivewidth}
\begin{document}
 
%float parameters
\renewcommand{\topfraction}{0.9}    % These three commands assure that floats
\renewcommand{\bottomfraction}{0.9} % (figures, tables) can cover a whole page
\renewcommand{\textfraction}{0.1}   % and no text is required
\renewcommand{\floatpagefraction}{0.8}
 
%define math alphabets for roman, boldface and teletype
\newcommand{\mrm}[1]{\mathrm{#1}}
\newcommand{\mbf}[1]{\mathbf{#1}}
\newcommand{\mtt}[1]{\mathtt{#1}}
\newcommand{\tsc}[1]{\textsc{#1}}
\newcommand{\tbf}[1]{\textbf{#1}}
\newcommand{\ttt}[1]{\texttt{#1}}
\newcommand{\br}[1]{\overline{#1}}
\newlength{\tmplen}
\newcommand{\clab}[1]{\tiny\settowidth{\tmplen}{\scriptsize#1}%
\colorbox{white}{\textcolor{white}{#1}}\hspace*{-1.27\tmplen}\scriptsize#1}

%some frequent math symbols
\def\lsim{\mathrel{\rlap{\lower4pt\hbox{\hskip1pt$\sim$}}
    \raise1pt\hbox{$<$}}}                % less than or approx. symbol
\def\gsim{\mathrel{\rlap{\lower4pt\hbox{\hskip1pt$\sim$}}
    \raise1pt\hbox{$>$}}}                % greater than or approx. symbol
\newcommand{\half}{$\frac{1}{2}$}        % 1/2.
 
%some frequent physics symbols
\newcommand{\alphas}{\alpha_{\mathrm{s}}}
\newcommand{\alphaem}{\alpha_{\mathrm{em}}}
\newcommand{\pT}{\ensuremath{p_{\perp}}}
\newcommand{\pTs}{\ensuremath{p^2_{\perp}}}
\newcommand{\kT}{\ensuremath{k_{\perp}}}
\newcommand{\pTmin}{p_{\perp\mathrm{min}}}
\newcommand{\pTmax}{p_{\perp\mathrm{max}}} % new
\newcommand{\pTp}{{p'}_{\perp}} % new
\newcommand{\pTo}{p_{\perp 0}}
\newcommand{\pTevol}{p_{\perp\mathrm{evol}}}
\newcommand{\pTPOW}{p_{\perp\mathrm{POWHEG}}}
\newcommand{\ECM}{E_{\mathrm{CM}}}
\newcommand{\mmin}{\mathrm{min}}
\newcommand{\mmax}{\mathrm{max}}
\newcommand{\MeV}{\ensuremath{\,\mathrm{MeV}}}
\newcommand{\GeV}{\ensuremath{\,\mathrm{GeV}}}
\newcommand{\TeV}{\ensuremath{\,\mathrm{TeV}}}
\newcommand{\fm}{\ensuremath{\,\mathrm{fm}}}
\newcommand{\mb}{\ensuremath{\,\mathrm{mb}}}
\newcommand{\rem}{\ensuremath{\mathrm{rem}}}
 
%roman names for particles in math mode
\renewcommand{\b}{\mathrm{b}}
\renewcommand{\c}{\mathrm{c}}
\renewcommand{\d}{\mathrm{d}}
\newcommand{\e}{\mathrm{e}}
\newcommand{\f}{\mathrm{f}}
\newcommand{\g}{\mathrm{g}}
\renewcommand{\j}{\mathrm{j}}
\newcommand{\J}{\mathrm{J}}
\newcommand{\hrm}{\mathrm{h}}
\newcommand{\n}{\mathrm{n}}
\newcommand{\p}{\mathrm{p}}
\newcommand{\q}{\mathrm{q}}
\newcommand{\s}{\mathrm{s}}
\renewcommand{\t}{\mathrm{t}}
\renewcommand{\u}{\mathrm{u}}
\newcommand{\A}{\mathrm{A}}
\newcommand{\D}{\mathrm{D}}
\renewcommand{\H}{\mathrm{H}}
\newcommand{\K}{\mathrm{K}}
\newcommand{\Q}{\mathrm{Q}}
\newcommand{\W}{\mathrm{W}}
\newcommand{\Zz}{\mathrm{Z^0}}
\newcommand{\Znz}{\mathrm{Z}}
\newcommand{\bbar}{\overline{\mathrm{b}}}
\newcommand{\cbar}{\overline{\mathrm{c}}}
\newcommand{\dbar}{\overline{\mathrm{d}}}
\newcommand{\fbar}{\overline{\mathrm{f}}}
\newcommand{\nbar}{\overline{\mathrm{n}}}
\newcommand{\pbar}{\overline{\mathrm{p}}}
\newcommand{\qbar}{\overline{\mathrm{q}}}
\newcommand{\sbar}{\overline{\mathrm{s}}}
\newcommand{\tbar}{\overline{\mathrm{t}}}
\newcommand{\ubar}{\overline{\mathrm{u}}}
\newcommand{\Bbar}{\overline{\mathrm{B}}}
\newcommand{\Dbar}{\overline{\mathrm{D}}}
\newcommand{\Qbar}{\overline{\mathrm{Q}}}
\newcommand{\qval}{\ensuremath{\q_{\mrm{v}}}}
\newcommand{\qsea}{\ensuremath{\q_{\mrm{s}}}}
\newcommand{\qcmp}{\ensuremath{\q_{\mrm{c}}}}
\newcommand{\val}{\ensuremath{{\mrm{v}}}}
\newcommand{\sea}{\ensuremath{{\mrm{s}}}}
\newcommand{\cmp}{\ensuremath{{\mrm{c}}}}
 
%roman names for supersymmetric particles in math mode
\newcommand{\sg}{\tilde{\mathrm{g}}}
\newcommand{\sq}{\tilde{\mathrm{q}}}
\newcommand{\sqd}{\tilde{\mathrm{d}}}
\newcommand{\squ}{\tilde{\mathrm{u}}}
\newcommand{\sqc}{\tilde{\mathrm{c}}}
\newcommand{\sqs}{\tilde{\mathrm{s}}}
\newcommand{\st}{\tilde{\mathrm{t}}}
\newcommand{\schi}{\tilde{\chi}}

%useful shorthands
\newcommand{\sigpp}{\sigma_{\mathrm{pp}}}
\newcommand{\nbphys}{\ensuremath{\langle n \rangle}}
\newcommand{\nb}{\ensuremath{\bar{n}}}
\newcommand{\nbb}{\ensuremath{\nb(b)}}
\newcommand{\nbbnz}{\ensuremath{\nbb \vert_{n \ne 0}}}
\newcommand{\ol}{\ensuremath{\tilde{\mathcal{O}}(b)}}
\newcommand{\olap}{\ensuremath{\tilde{\mathcal{O}}(b, x_1, x_2)}}
\newcommand{\omax}{\ensuremath{\tilde{\mathcal{O}}_{\mrm{max}}(b)}}
\newcommand{\ntb}{\ensuremath{\bar{n}(b)}}
\newcommand{\pint}{\ensuremath{P_{\mrm{int}}(b)}}
\newcommand{\sh}{\ensuremath{\sigma_{\mrm{hard}}}}
\newcommand{\snd}{\ensuremath{\sigma_{\mrm{ND}}}}
\newcommand{\stot}{\ensuremath{\sigma_{\mrm{tot}}}}
 
%new list environments to replace itemize and enumerate
\newenvironment{Itemize}{\begin{list}{$\bullet$}%
{\setlength{\topsep}{0.2mm}\setlength{\partopsep}{0.2mm}%
\setlength{\itemsep}{0.2mm}\setlength{\parsep}{0.2mm}}}%
{\end{list}}
\newcounter{enumct}
\newenvironment{Enumerate}{\begin{list}{\arabic{enumct}.}%
{\usecounter{enumct}\setlength{\topsep}{0.2mm}%
\setlength{\partopsep}{0.2mm}\setlength{\itemsep}{0.2mm}%
\setlength{\parsep}{0.2mm}}}{\end{list}}

\newcommand{\herwig}{HERWIG\xspace}
\newcommand{\herpp}{Herwig++\xspace}
\newcommand{\phojet}{\textsc{Phojet}\xspace}
\newcommand{\jimmy}{\textsc{Jimmy}\xspace}
 
%set sloppy attitude to line breaks
\sloppy
 
\pagestyle{empty}
 
\begin{flushright}
LU TP 11-06\\
MCnet/11/03\\
January 2011
\end{flushright}
 
\vspace{\fill}
 
\begin{center}
{\LARGE\bf Multiparton Interactions\\[2mm]
with an $x$-dependent Proton Size}\\[10mm]
{\Large R.~Corke\footnote{richard.corke@thep.lu.se} and %
T.~Sj\"ostrand\footnote{torbjorn@thep.lu.se}} \\[3mm]
{\it Theoretical High Energy Physics,}\\[1mm]
{\it Department of Astronomy and Theoretical Physics,}\\[1mm]
{\it Lund University,}\\[1mm]
{\it S\"olvegatan 14A,}\\[1mm]
{\it S-223 62 Lund, Sweden}
\end{center}
 
\vspace{\fill}
 
\begin{center}
{\bf Abstract}\\[2ex]
\begin{minipage}{\abstwidth}
Theoretical arguments, supported by other indirect evidence, suggest that
the wave function of high-$x$ partons should be narrower than that of
low-$x$ ones. In this article, we present a modification to the variable
impact parameter framework of \tsc{Pythia 8} to model this effect. In
particular, a Gaussian hadronic matter profile is introduced, with a
width dependent on the $x$ value of the constituent being probed.
Results are compared against the default single- and double-Gaussian
profiles, as well as an intermediate overlap function.
\end{minipage}
\end{center}
 
\vspace{\fill}
 
\clearpage
\pagestyle{plain}
\setcounter{page}{1}

\section{Introduction}
\label{sec:intro}

Multiparton interactions (MPI) are an unavoidable consequence of colliding
hadrons at high energies. The theoretical description of soft MPI poses
particular challenges, due to the limited understanding of non-perturbative
QCD. Unfortunately, it is exactly this physics which is vital for
describing minimum-bias (MB) and underlying events (UE) at hadron
colliders, such as the LHC. To this end, phenomenological models are often
introduced, to provide the best description possible.

The original MPI model, introduced in earlier versions of \tsc{Pythia},
extends the perturbative picture down to very low $\pT$ scales, such that
one can view all events as containing one or more interactions
\cite{Sjostrand:1987su,Bengtsson:1987kr,Sjostrand:2006za}.
These low-$\pT$ interactions fill the role of cut Pomerons
\cite{Amati:1962nv,Abramovskii:1972zz}, stretching
colour fields longitudinally across an event, which later fragment. Of
course, these colour fields can also stretch to higher-$\pT$ partons,
giving a smooth transition to (mini)jets, and a unified picture of MB and
UE physics. This model has been updated in recent times, and forms a
part of the interleaved parton shower and MPI framework of \tsc{Pythia 8}
\cite{Sjostrand:2004pf,Sjostrand:2004ef,Sjostrand:2007gs,Corke:2010yf}.

Many other models for the structure of hadronic events have been 
formulated, that are all based on some kind of multiple interactions 
framework, be it in the form of soft or (semi)hard interactions, or a 
mixture thereof \cite{Alessandro:2011wt}. A few implementations are 
formulated with a view to be used also for hard-scale physics within 
and beyond the Standard Model \cite{Buckley:2011ms}, such as 
\tsc{Herwig} \cite{Marchesini:1988hj,Marchesini:1991ch, 
Butterworth:1996zw,Borozan:2002fk,Bahr:2008dy,Bahr:2008pv} and 
\tsc{Sherpa} \cite{Gleisberg:2008ta,Ryskin:2009tj}. Others put more 
emphasis on the soft physics aspects, including the relations between 
elastic, diffractive and non-diffractive topologies, using concepts 
such as Dual Topological Unitarization \cite{Capella:1992yb} and 
Reggeon Field Theory \cite{Gribov:1968fc}. Examples thereof include
\tsc{Phojet} \cite{Engel:1994vs,Engel:1995yda}, \tsc{Dpmjet} 
\cite{Bopp:2005cr}, \tsc{Epos} \cite{Werner:2005jf}, \tsc{Sibyll} 
\cite{Ahn:2009wx}, and \tsc{Qgsjet} \cite{Ostapchenko:2010vb}.

In all of these programs the proton is handled as an extended object.
That way an eikonal description \cite{Glauber:1959xx} can be used, 
wherein the probability for an event to be produced is largest 
for head-on collisions and decreases for increasing impact parameter.
The standard assumption for most of these scenarios is that the 
partons are distributed inside the protons according to a Gaussian,
with the same radius for all parton species and momenta. 
There is no specific reason for this ansatz, but it makes for   
simple algebra in going from a three-dimensional spherical ansatz
to a two-dimensional impact-parameter plane, and for convoluting 
these distributions for the two colliding hadrons. Other shapes have 
been used, e.g.\ the electromagnetic form factor in \tsc{Herwig},
and some \tsc{Pythia} alternatives to be described later. 
A collision-energy-dependent radius is often used, and sometimes 
two different radii for soft and hard interactions, but these 
possibilities still offer fairly little flexibility. The one notable 
exception we are aware of is \tsc{Dipsy}
\cite{Avsar:2005iz,Avsar:2006jy,Avsar:2007xg}, see further below. 

The key objective of the current article will thus be to study the
consequences if one of the conventional constraints is relaxed,
namely that high- and low-momentum partons have the same 
impact-parameter profile. This should actually be considered as 
the expected behaviour, rather than an exotic variant, as follows. 

The size of the proton is finite, owing to confinement, 
but exactly how it should be defined is ambiguous. Low-energy 
measurements give a root-mean-squared (RMS) charge radius $\approx 0.88$~fm
\cite{Mohr:2008fa}. Combined with the mass gap of QCD --- the lightest free
state being the pion --- this leads to a finite proton--proton
strong-interaction cross section $\sigpp$ (while the electromagnetic one is
infinite, the photon being massless).

This cross section can vary as a function of energy, but its
growth is limited by the Froissart--Martin 
bound \cite{Froissart:1961ux, Martin:1965jj}. The intuitive 
idea underlying this bound is that the pion Yukawa potential 
fall-off $g e^{-m_{\pi}r}/r$ sets the maximum impact parameter 
$b_{\mathrm{max}}$ of interactions to be roughly where 
$|g| \exp(-m_{\pi}b_{\mathrm{max}}) = 1$, i.e.\
$\sigma \simeq \pi b_{\mathrm{max}}^2  \simeq (\pi/m_{\pi}^2) \ln^2|g|$.
Since it can be shown that $|g|$ can increase at most like 
a power of the collision energy under general conditions which 
should hold for QCD, it follows that $\sigma \propto \ln^2s$
provides an upper bound. The numerical prefactor to the bound
\cite{Lukaszuk:1967zz} is far from saturated at current energies, 
however, and work to improve on it is ongoing 
\cite{Martin:2009pt,Wu:2010sd}. 

The experimental observation of an increasing total cross section 
is reinforced by studies of the differential elastic cross section
\cite{Chou:1968bc,Cheng:1974ty}, from which it is concluded that 
the proton gets ``blacker, edgier, larger'' with increasing energy 
\cite{Henzi:1984fz,Bourrely:1984gi}.

By Gribov theory, the high-$s$ behaviour can be related to a 
low-$x$ one, with the size of the proton growing proportionally
to $\ln(1/x)$. Qualitatively (but not quantitatively, see below)
this can be understood as a transverse random walk
\cite{Frankfurt:2005mc} in a BFKL \cite{Kuraev:1977fs,Balitsky:1978ic} 
evolution, where a few initial high-$x$ partons fairly close to 
the center of the proton emit a cascade of partons towards 
lower $x$ scales, and in the process these partons diffuse to be 
spread over a larger area. A more formal definition
can be obtained by the Balitsky-JIMWLK evolution equations for
hadronic amplitudes \cite{Balitsky:1995ub}, which also can be 
described by the Color Glass Condensate formalism \cite{Hatta:2005rn}. 

Mueller's dipole cascade model \cite{Mueller:1993rr,Mueller:1994jq} 
offers a formulation of the BFKL evolution in transverse coordinate
space, and so gives direct access to information on the spread of 
partons at different $x$ scales. The \tsc{Dipsy} generator provides
a complete implementation, where effects of energy--momentum 
conservation, saturation, gluon recombination and the running of 
$\alphas$ are consistently taken into account. 
One important message that comes out of the numerical studies is 
that the Froissart--Martin bound is violated asymptotically
in the evolution equations, unless confinement is also built into 
the gluon propagator, in which case the $\ln^2s$ behaviour is 
nicely obtained \cite{Avsar:2008dn}. 

The \tsc{Dipsy} generator can also be used to study a number of 
further issues, such as diffraction \cite{Flensburg:2010kq} and
elliptic flow \cite{Avsar:2010rf}. So far it has not been used for
comparisons with MB event properties at hadron colliders,
however, and is not well suited for UE studies.

Generalized parton distributions offer an alternative approach 
to explore the transverse size of the proton \cite{Burkardt:2002hr},
and to understand some MPI phenomenology \cite{Diehl:2010dr}.
Again an $x$ dependence is obtained, where the proton radius 
vanishes in the limit $x \to 1$ (in part a natural consequence
of a center-of-gravity definition of the origin). 

In the current article we will not attempt to trace the evolution
of cascades in $x$. Rather we will assume that the impact-parameter
distribution of partons at any $x$ can be described by a simple 
Gaussian, $\exp(-b^2/a^2(x))$, with a width that grows like
$a(x) = a_0 \left( 1 + a_1 \ln(1/x) \right)$.
The coefficients $a_0$ and $a_1$ are tuned to the parameterised 
shape of $\sigpp(s)$ in the following. The potential overlap between 
two protons will be described only in terms of their size
at their respective $x$ values. In principle one should also 
include a third scale, related to the transverse distance the 
exchanged propagator particle, normally a gluon, could travel. 
This distance should be made dependent on the $\pT$ scale of the 
interaction. For simplicity we will not consider this further 
complication here, and only take the propagator distance into account  
by allowing a finite effective radius also for $x \to 1$.
For minimum-bias studies, the results will be rather insensitive to the
behaviour at large $x$ since, at high energies, the bulk of MPI occurs at
small $x$. This choice will play more of a role in the underlying event of
hard processes, but is not studied further.

In Section~\ref{sec:mpi}, some relevant aspects of the existing MPI model
are given, before the modified impact parameter framework is introduced.
Some results are shown in Section~\ref{sec:results}, both in comparison to
other matter profiles and to data, before a summary and
outlook is given in Section~\ref{sec:summary}.

\section{Multiparton interaction framework}
\label{sec:mpi}
The starting point is the hadronic perturbative cross section
\begin{equation}
 \frac{\d\sigma}{\d\pTs} = \sum_{i,j} \iint \d x_1 \, \d x_2 \,
 f_i(x_1, Q^2) \, f_j(x_2, Q^2) \, \frac{\d \hat{\sigma}}{\d\pTs} ~,
\label{eq:dsigmadpt}
\end{equation} 
where $\d \hat{\sigma} / \d\pTs$ gives the partonic QCD $2 \to 2$ cross
section, and $f_i$ and $f_j$ the PDF factors of the two incoming hadrons.
In the modelling of soft MPI activity, there are two key observations that
can be made. First, the QCD $2 \to 2$ cross section contains a $1/\pT^4$
divergence in the $\pT \to 0$ limit. Second, the total integrated cross
section down to some low-$\pT$ limit
\begin{equation}
 \sh(\pTmin) = \int^{s/4}_{\pTmin^2} \frac{\d\sigma}{\d\pTs} ~ \d\pTs ~,
\end{equation}
becomes comparable to the total cross section for $\pTmin \approx 2-5\GeV$ at
current collider energies.

The original MPI model addresses these issues as follows
\cite{Sjostrand:1987su}. It is observed that $\sh$ gives the hadron--hadron
cross section and not the parton--parton one. If, in one hadron collision,
many parton--parton interactions are possible, then
$\nbphys(\pTmin) = \sh(\pTmin) / \stot$
gives the average number of parton-parton scatterings above $\pTmin$ per
event. In dealing with non-diffractive inelastic events only, as in this
article, the cross section for hard interactions, $\sh(\pTmin)$, must be
distributed among the $\snd$ events, such that
$\nbphys(\pTmin) = \sh(\pTmin) / \snd$. 

This is still not a solution to the divergence of the cross section in the
$\pT \to 0$ limit. The average $\hat{s}$ of scatterings decreases slower
with $\pTmin$ than the number of interactions increases, which would lead
to an infinite amount of scattered partonic energy. One part of the
solution is the need to include longitudinal correlations, including energy
and momentum conservation effects. In the most recent iterations of the
model, this is handled using a model dependent PDF rescaling procedure
\cite{Sjostrand:2004pf}.

This effect alone is too weak, however, and the model additionally
introduces the idea of colour screening to regularise the $\pT \to 0$
divergence. The concept of a perturbative cross section is based on the
assumption of free incoming states, which is not the case when partons are
confined in colour-singlet hadrons. One therefore expects a colour charge
to be screened by the presence of nearby anti-charges; that is, if the
typical charge separation is $d$, gluons with a transverse wavelength $\sim
1 / \pT > d$ are no longer able to resolve charges individually, leading to
a reduced effective coupling. This is introduced by reweighting the
interaction cross section such that it is regularised according to
\begin{equation}
\frac{\d \hat{\sigma}}{\d \pT^2} \propto
\frac{\alphas^2(\pT^2)}{\pT^4} \rightarrow
\frac{\alphas^2({\pT^2}_0 + \pT^2)}{({\pT^2}_0 + \pT^2)^2}
,
\label{eqn:pt0}
\end{equation}
where $\pTo$ (related to $1 / d$ above) is now a free parameter in the
model.

This parameter has an energy dependence, and the ansatz used is that it
scales in a similar manner to the total cross section, i.e.\ driven by an
effective power related to the Pomeron intercept \cite{Donnachie:1992ny},
which in turn could be related to the small-$x$ behaviour of parton
densities. This leads to a scaling
\begin{equation}
\pTo(E_{\mathrm{CM}}) = p_{\perp0}^{\mathrm{ref}} \times \left(
\frac{E_{\mathrm{CM}}}{E_{\mathrm{CM}}^{\mathrm{ref}}}\right)%
^{E_{\mathrm{CM}}^{\mathrm{pow}}} ~,
\label{eq:pT0scaling}
\end{equation}
where $E_{\mathrm{CM}}^{\mathrm{ref}}$ is some convenient reference energy
and $p_{\perp0}^{\mathrm{ref}}$ and $E_{\mathrm{CM}}^{\mathrm{pow}}$ are
parameters to be tuned to data.

\subsection{Hadronic matter distribution}
\label{sec:hmd}

In the original MPI framework of \cite{Sjostrand:1987su}, events are
characterised by a varying impact parameter, $b$, representing a classical
distance of closest approach between the two incoming hadrons.
The hadronic matter is assumed to have a spherically symmetric
distribution, taken to be the same for all parton species and momenta.
The time-integrated overlap between the two incoming matter distributions
at an impact parameter, $b$, is given by
\begin{equation}
  \ol = \int \d t \int \d^3x ~
        \rho (x, y, z) ~
        \rho (x, y, z - \sqrt{b^2 + t^2}) ~,
\end{equation}
where the $\rho$'s give the matter distributions after a scale change to take
into account the boosted nature of the hadrons. There are currently three
different matter profiles available:
\begin{itemize}
 \item[1)] Single Gaussian: a simple Gaussian with no free parameters
 \begin{equation}
   \rho(r) \propto \exp (-r^2) ~.
 \end{equation}

 \item[2)] Double Gaussian: a core region, radius $a_2$, contains a
           fraction $\beta$ of the total hadronic matter, embedded in a
           larger hadron of radius $a_1$. The default parameters for this
           profile are $a_2 / a_1 = 0.4$ and $\beta = 0.5$
 \begin{equation}
  \rho(r) \propto (1 - \beta) ~ \frac{1}{a_1^3} \,
                  \exp \left( - \frac{r^2}{a_1^2} \right) +
                  \beta ~ \frac{1}{a_2^3} \,
                  \exp \left( - \frac{r^2}{a_2^2} \right) ~.
 \end{equation}

 \item[3)] Overlap function: $\ol$, rather than $\rho(r)$, is
           parameterised by a single parameter, $p$. When $p = 2$, this
           gives the single Gaussian behaviour, while when $p = 1$, results
           are similar to the default double Gaussian behaviour
 \begin{equation}
  \ol \propto \exp \left( -b^{p} \right) ~.
 \end{equation}
\end{itemize}

In what follows, we relax the assumption that this distribution remains the
same for all momenta, such that the wavefunction for small-$x$ partons is
broader in spatial extent than for large-$x$ ones. In particular, a
form
\begin{equation}
  \rho(r, x) \propto \frac{1}{a^3(x)} \, \exp
                     \left( - \frac{r^2}{a^2(x)} \right) ~, 
  \label{eq:loggauss}
\end{equation}
\begin{equation}
  a(x) = a_0 \left( 1 + a_1 \ln \frac{1}{x} \right) ~,
  \label{eq:ax}
\end{equation}
is chosen, where $x$ represents the momentum fraction of the parton being
probed within the hadron, $a_0$ is a constant to be tuned according to the
non-diffractive cross section (detailed below) and $a_1$ is a free
parameter. When $a_1 = 0$, the single Gaussian profile is recovered. With
this matter profile, the time-integrated overlap is given by
\begin{equation}
  \olap =
  \frac{1}{\pi}
  \frac{1}{a^2(x_1) + a^2(x_2)} \,
  \exp \left( - \frac{b^2}{a^2(x_1) + a^2(x_2)} \right) ~,
  \label{eq:olap}
\end{equation}
where the normalisation has been chosen such that
\begin{equation}
  \int \olap ~ \d^2b = 1 ~.
  \label{eq:normcond}
\end{equation}

\subsection{Impact parameter framework}
\label{sec:ipf}
Within the framework, the number of interactions is assumed to be
distributed according to a Poissonian distribution. If $\ntb$ gives the
average number of interactions when two hadrons pass each other with an
impact parameter $b$, the probability that there is at least one
interaction is given by
\begin{equation}
  \pint = 1 - e^{- \nbb} ~.
\end{equation}
This gives the requirement for an event to be produced in the first place.
The average number of interactions per event at impact parameter $b$ is
therefore given by
\begin{equation}
  \nbbnz = \frac{\nbb}{\pint} ~.
\end{equation}
When integrated over all impact parameters, the relation
$\nbphys = \sh / \snd$ (Sec.~\ref{sec:mpi}) must still hold, giving
\begin{equation}
  \nbphys =
  \frac{\int \nbbnz \, \pint \, \d^2b}{\int \pint \, \d^2b} =
  \frac{\int \nbb \, \d^2b}{\int \left( 1 - e^{- \nbb} \right) \d^2b} =
  \frac{\sh}{\snd} ~.
  \label{eq:navg}
\end{equation}
Defining the shorthand $X = (x_1, \, x_2, \, \pTs)$ and
$\d X = \d x_1 \, \d x_2 \, \d \pTs$, $\sh$ may now be written as
\begin{equation}
  \sh = \int \d X \frac{\d \sigma}{\d X} 
      = \iint \d X \, \d^2b \,
        \frac{\d \sigma}{\d X} \, \olap ~,
\end{equation}
where eq.~(\ref{eq:normcond}) has been used to associate an
impact-parameter profile with each $X$ coordinate. Here, $\d \sigma / \d X$
gives the convolution of PDF factors and the (regularised) hard partonic
cross section
\begin{equation}
  \frac{\d \sigma}{\d X} =
  f_1(x_1, \pTs) \, f_2(x_2, \pTs) \,
  \left. \frac{\d \hat{\sigma}}{\d \pTs} \right|_{\mrm{reg}} ~.
  \label{eq:pTall}
\end{equation}
Comparing with eq.~(\ref{eq:navg}), this gives the average number of
interactions at an impact parameter $b$ to be
\begin{equation}
  \nbb = \int \d X \, \frac{\d \sigma}{\d X} \,
         \olap ~.
  \label{eq:nbb}
\end{equation}

One can now give a geometrical interpretation to $\sh$ and $\snd$
\begin{eqnarray}
  \sh &=& \int \nbb \, \d^2 b ~,
  \label{eq:sigmahard}\\
  \snd &=& \int \pint \, \d^2b =
  \int \left( 1 - e^{- \nbb} \right) \d^2b ~,
  \label{eq:sigmand}
\end{eqnarray}
such that eq.~(\ref{eq:navg}) is fulfilled. This determines the value of
$a_0$ as follows. Eq.~(\ref{eq:sigmahard}) fixes the total area of $\nbb$,
within the constraint that it is possible to have either a large width
$a_0$ and a small height $\nb(0)$, or the other way around. In the former
case, $1 - \exp(- \nb(0)) \approx \nb(0)$, giving $\snd \approx \sh$. In the
latter, strong saturation effects lead to
$1 - \exp(- \nb(0)) \ll \nb(0)$, giving $\snd \ll \sh$. The saturation
corrections increase monotonically with $\nb(0)$ and so a unique solution
for $a_0$ is defined. This is studied further in Sec.~\ref{sec:psize}.

\subsection{Impact parameter selection}
\label{sec:bsel}

In picking the hardest interaction in an event, $p_{\perp 1}^2$, the naive
probability for a collision must be multiplied by the probability that
there were no harder ones at scales $\pTs > p_{\perp 1}^2$. Using the
notation
\begin{equation}
  \nbb = \int \frac{\d \nbb}{\d \pTs} \, \d \pTs ~,
\end{equation}
the total probability distribution is now
\begin{equation}
  \frac{\d P_{\mrm{hardest}}}{\d^2b \, \d p_{\perp 1}^2} =
  \frac{\d \nbb}{\d p_{\perp 1}^2} \,
  \exp \left( - \int^{s/4}_{p_{\perp 1}^2}
  \frac{\d \nbb}{\d \pTs} \, \d \pTs \right) ~.
  \label{eq:softpT}
\end{equation}

One possible way of generating events according to this distribution is
through trial interactions, similar to e.g. trial showers in CKKW-L
\cite{Lonnblad:2001iq}, in the following way. If the evaluation of the
Sudakov factor is temporarily deferred, then
\begin{equation}
  \frac{\d P_{\mrm{hardest}}}{{\d {\pTs}_1}} =
  \int \d^2b \, \frac{\d \nbb}{\d p_{\perp 1}^2} =
  \iint \d x_1 \, \d x_2 \, f_1(x_1, {\pTs}_1) \, f_2(x_2, {\pTs}_1) \,
  \left. \frac{\d \hat{\sigma}}{\d {\pTs}_1} \right|_{\mrm{reg}} ~.
  \label{eq:hardpT}
\end{equation}
${\pTs}_1$, $x_1$ and $x_2$ may then be picked according to the above
distribution, before an impact parameter $b$ is selected according to
$\olap \, \d^2b$. The scale of a trial MPI interaction, ${\pTs}_2$, may
then be generated for this $b$ value, as described in the next section. It
is important to note that the ${\pTs}_2$ evolution is started from the
kinematical limit, $s / 4$, as for an event with no previous interaction.
If ${\pTs}_2 < {\pTs}_1$, then ${\pTs}_1$ and $b$ are accepted, else the
selection procedure must restart from the beginning.

The above provides a prescription for generating an inclusive sample of
non-diffractive events (hereafter referred to as minimum bias), but can
also be used to generate the MPI activity accompanying a pre-given hard
process. This is simplest in the case where the hard process in question is
already part of the set of processes contained in $\d \sigma / \d X$.
Here, ${\pTs}_1$ is provided by this hard process, and the MPI
framework should not generate any interactions at higher scales, or else
one would double count. Given a hard process at a scale ${\pTs}_1$, 
$b$ can be selected from $\olap \, \d^2b$, and then
retained with a probability equal to the Sudakov of eq.~(\ref{eq:softpT}).
Again, trial interactions are a possible way to generate this Sudakov
factor.

When the hard process is not contained in $\d \sigma / \d X$,
such as $\Zz$ production, the MPI framework can begin evolution at the
kinematical limit without any risk of double counting. For this discussion,
noting that $\pT$ is intended as a measure of hardness, we assume a scale
such as $\hat{s}$ to be a reasonable choice for this process. One choice
that must be made relates to which interaction is used in selecting the
impact parameter for an event. If we decide that it is always the hardest
interaction in an event, it should be remembered that there is now the
possibility that this is an MPI, although this is rather unlikely for a
process already picked to be hard.

When the pre-given hard process has a scale above $10 - 20 \GeV$, the
Sudakov of eq.~(\ref{eq:softpT}) will be close to unity, meaning that $b$
can be directly selected from $\olap \, \d^2b$ and any ambiguity will be
minor. For hard processes around these scales, the correct procedure is
less clear. One choice would be to retain the selection according to $\olap
\, \d^2b$ only, while another would be to additionally apply a Sudakov
weight to the selection.

In the hard process studies that follow, the impact parameter is always
selected according to the hard process (not necessarily the hardest
interaction, as above), and this selection is not weighted with a Sudakov.
As above, for hard processes above $10 - 20 \GeV$, these choices do not
greatly affect the outcome. With the single Gaussian, double Gaussian and
overlap matter profiles, the MPI framework will give exactly the same
impact parameter profile for e.g. a $1\TeV$ $Z'$ as for $\Zz$ production.
The amount of MPI activity is modified by longitudinal correlations,
however, such that the $1\TeV$ $Z'$ will, on average, have less activity.
The new matter profile, with its varying width, instead, dynamically
changes to give more underlying activity for the more massive state, given
the differing $x$ values that enter.

\subsection{Subsequent evolution}

Once $b$ has been fixed, as in the previous section, the remaining sequence
of multiple interactions must be generated according to
\begin{equation}
  \frac{\d P}{\d p_{\perp i}} =
  \frac{\d \nbb}{\d p_{\perp i}} \,
  \exp \left( - \int^{p_{\perp i - 1}}_{p_{\perp i}}
  \frac{\d \nbb}{\d \pTs} \, \d \pTs \right) ~.
  \label{eq:nextpT}
\end{equation}
This can be achieved through the veto algorithm \cite{Sjostrand:2006za}, as
follows. Temporarily neglecting the impact parameter dependence, an
overestimate of the form
\begin{equation}
  \iint \d x_1 \, \d x_2 \, f_1(x_1, \pTs) \, f_2(x_2, \pTs) \,
  \left. \frac{\d \hat{\sigma}}{\d \pTs} \right|_{\mrm{reg}} \le
  \frac{N}{(\pTs + r \, p_{\perp 0}^2)^2}
\end{equation}
can be used, where $r$ and $N$ are tunable factors; the former to help
flatten the correction ratio, to improve generation efficiency, and the
latter to ensure that the overestimate sits above the cross section over
the entire phase space. With the impact parameter dependence present, an
additional factor, giving the maximum of the overlap distribution is
introduced
\begin{equation}
  \olap \le \omax = \frac{1}{2 \pi a_0^2} \,
  \exp \left( - \frac{b^2}{2 a_{\mrm{max}}^2 } \right) ~,
\end{equation}
\begin{equation}
  a_{\mrm{max}} = a_0 \left( 1 + a_1 \ln \frac{1}{x_{\mrm{min}}} \right) ~,
\end{equation}
where the first factor gives the maximum height of the distribution, while
the exponential width is dictated by smallest $x$ values reached. This then
gives a total overestimate
\begin{equation}
  \frac{\d \nbb}{\d \pTs} \le
  \frac{N \, \omax}{(\pTs + r \, p_{\perp 0}^2)^2} ~,
  \label{eq:overestimate}
\end{equation}
giving a uniform overestimation for all $x_1$ and $x_2$
\begin{equation}
  \int \omax \, \d^2b = \frac{a_{\mrm{max}}^2}{a_0^2} ~.
\end{equation}
Eq.~(\ref{eq:overestimate}) is inserted into eq.~(\ref{eq:nextpT}) to pick
the next $\pT$ scale. The additional acceptance weight for this interaction
is now given by
\begin{equation}
  \frac{\olap}{\omax} = 
  \frac{2 a_0^2}{a^2(x_1) + a^2(x_2)} \,
  \exp \left( \frac{b^2}{2 a_{\mrm{max}}^2} -
              \frac{b^2}{a^2(x_1) + a^2(x_2)} \right) ~.
\end{equation}
In case of failure, the evolution in $\pT$ is continued downwards from the
rejected $\pT$ value.

\section{Results}
\label{sec:results}

\subsection{Growth of the total cross section}
\label{sec:psize}
In principle, $a_1$, as introduced so far, is a free parameter. If,
however, as suggested earlier, the wider profile of low-$x$ partons
is to account for the growth of the total cross section (or the inelastic
non-diffractive one, as in this model), then it can be constrained by the
requirement that $a_0$ should be independent of energy.

\begin{figure}
\centering
\includegraphics[scale=0.63,angle=270]{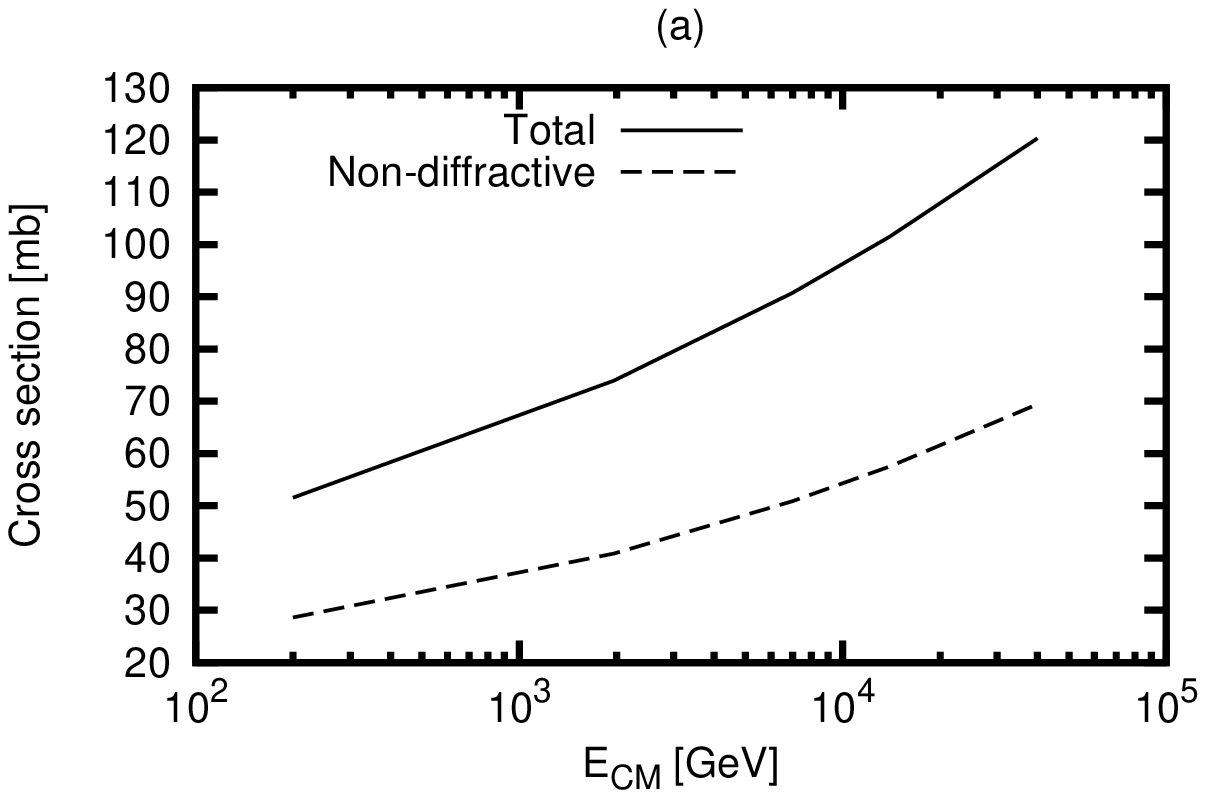}
\includegraphics[scale=0.63,angle=270]{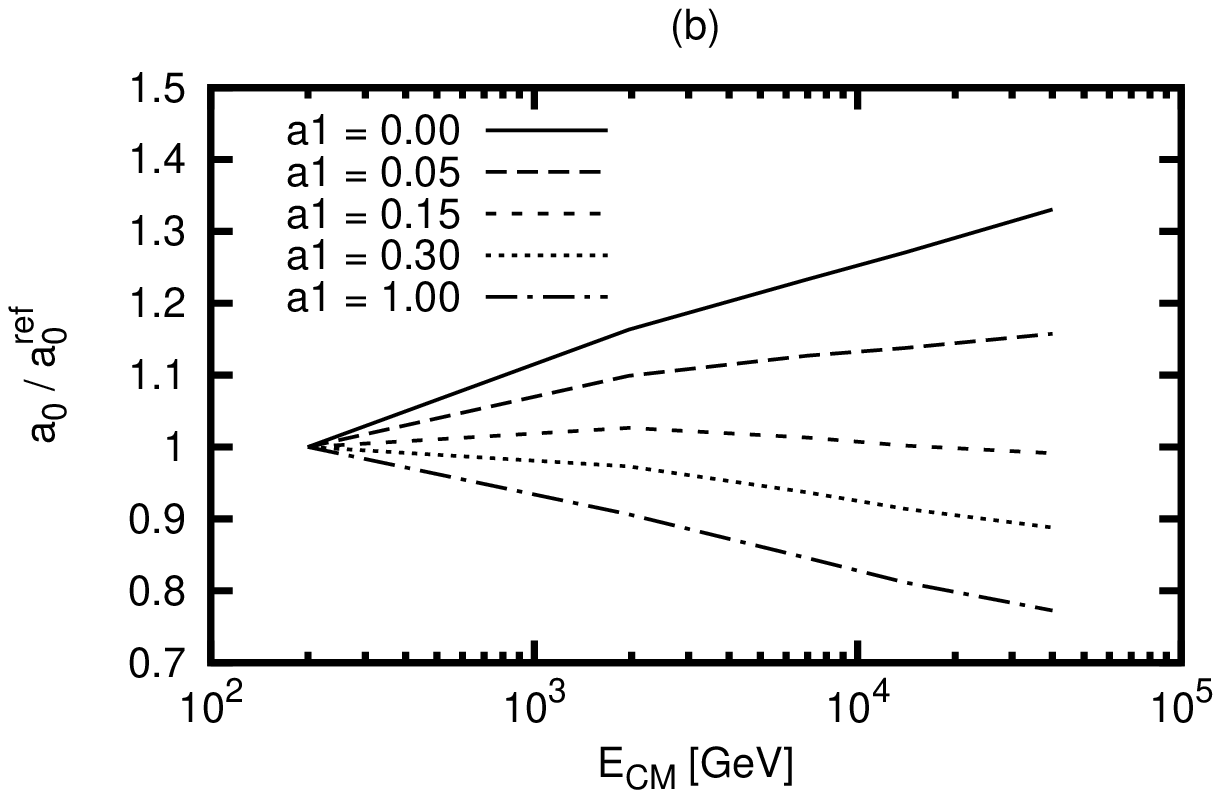}
\caption{(a) The rise of the total and non-diffractive $\p\p$ cross section
with energy, and (b) the ratio $a_0(E_{\mrm{CM}}) / a_0(200\GeV)$, over the
same energy range, for a set of different $a_1$ values
\label{fig:xsec}}
\end{figure}

The total cross section is taken from a Donnachie-Landshoff
parameterisation \cite{Donnachie:1992ny}. It is also necessary to break
this down into elastic, diffractive and non-diffractive components, which
is done based on a parameterisation incorporating empirical corrections
such that the elastic and diffractive cross sections do not exceed the
total at higher energies \cite{Schuler:1993wr}. In all that follows, we will
deal explicitly with $\p\p$ collisions. The assumed rise of the total and
non-diffractive cross sections are shown as a function of centre-of-mass
energy in Fig.~\ref{fig:xsec}a.

In the calculation of $a_0$, $\d \sigma / \d X$ enters, giving a dependence
on PDFs and the $\pTo$ used to regularise the cross section. In all that
follows, the parameters of Tune 4C \cite{Corke:2010yf}, a tune to early LHC
data, are used. It is the relative variation of $a_0$ as a function of
energy that is of interest here, and in Fig.~\ref{fig:xsec}b the ratio
$a_0(E_{\mrm{CM}}) / a_0(200\GeV)$ is shown over the same range of
energies, for a set of different $a_1$ values. A value of $a_1 = 0.15$
gives an $a_0$ that is relatively stable across this energy range.

\begin{figure}
\centering
\includegraphics[scale=0.63,angle=270]{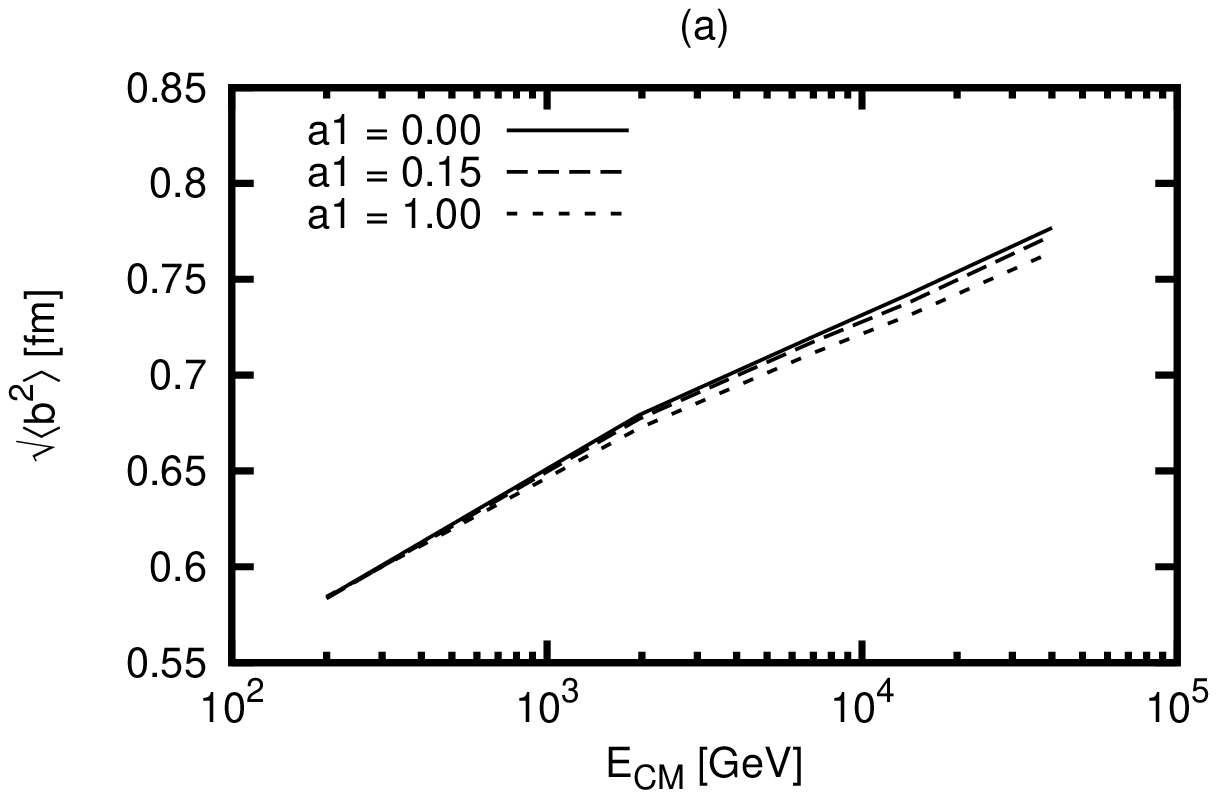}
\includegraphics[scale=0.63,angle=270]{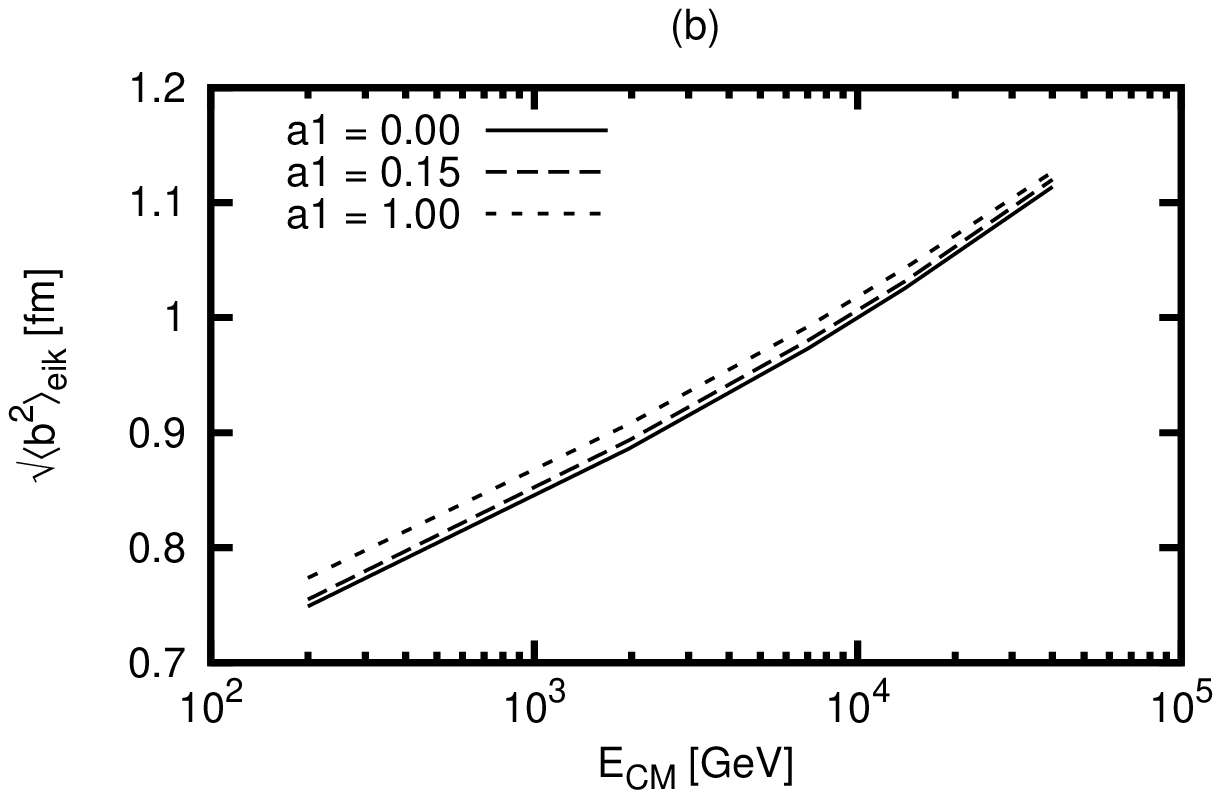}
\caption{Root-mean-squared value of (a) $\nb$ and (b) $P_\mrm{int}$ as a
function of the centre-of-mass energy
\label{fig:psize}}
\end{figure}

As discussed in Sec.~\ref{sec:ipf}, through the eikonalisation procedure
and tuning of the $a_0$ parameter, the width of the matter profile has an
absolute meaning, related to the size of the incoming hadron.
As $a_1$ is increased from zero, $\nbb$, after integration over $x$ values,
is higher both at small and large $b$ values, and smaller at intermediate
$b$ values, such that the total area is conserved. In some ways the shape
is similar to what one would expect from a double Gaussian matter profile,
where the central core of matter would tend to push the distribution up at
small $b$, while the tail would also have a larger content, due to the
peripheral component.

Before eikonalisation, and given that the form of eq.~(\ref{eq:loggauss})
stretches out to infinity, a simple measure is given by the
RMS value of $\nbb$
\begin{equation}
  \langle b^2 \rangle =
  \frac{\int b^2 \, \nbb \, \d^2b}{\int \nbb \, \d^2b}
  = \frac{1}{\sh} \int b^2 \, \nbb \, \d^2b ~.
\end{equation}
The variation of $\langle b^2 \rangle$ with energy is shown in
Fig.~\ref{fig:psize}a. The slope slightly decreases with the rise of
$a_1$, dependent on the enhancement of $\nbb$ at low $b$ relative to high.

Assuming a single Gaussian matter profile, the
width, $a$, needed to give the same $\langle b^2 \rangle$ is given by
$a = \sqrt{3 \langle b^2 \rangle} / 2$, noting that the RMS radius is
defined in 3 rather than 2 dimensions. This gives values beneath the
conventional charge RMS radius. We do not study this further here, but do
note that the eikonalisation procedure used ignores any contribution from
diffractive components, also noted below.

The same quantity, after the eikonalisation procedure can also be obtained
from
\begin{equation}
  \langle b^2 \rangle_{\mrm{eik}} =
  \frac{\int b^2 \, \pint \, \d^2b}{\int \pint \, \d^2b}
  = \frac{1}{\snd} \int b^2 \, \pint \, \d^2b ~.
\end{equation}
The variation of this quantity with energy is shown in
Fig.~\ref{fig:psize}b. After eikonalisation, the component of $\nbb$ at
large $b$ becomes more important, leading to an increase as $a_1$ grows
from zero, due to the contribution of low-$x$ partons.

A final consistency check is provided by the standard eikonal formulae
\cite{Glauber:1959xx}, providing a relation between the total and inelastic
cross sections
\begin{equation}
  \sigma_{\mrm{inel}} = \int \d^2 b \, \left( 1 - e^{2\chi(b)} \right) ~,
  \label{eq:eikinel}
\end{equation}
\begin{equation}
  \sigma_{\mrm{tot}} = 2 \int \d^2 b \, \left( 1 - e^{\chi(b)} \right) ~.
  \label{eq:eiktot}
\end{equation}
From the former equation, we can identify the eikonal function
$\chi(b) = \nbb / 2$. Using the latter equation to
calculate the total cross section, the result is consistently below the
total cross section of Fig.~\ref{fig:xsec}a by around $10 - 20\%$. As noted
above, the diffractive component has been ignored in the above framework,
and is a potential source for these deviations, including the low
$\langle b^2 \rangle$ values noted previously.

\subsection{Hard processes}

In what follows, comparisons are made between the different hadronic matter
distributions (Sec.~\ref{sec:hmd}). For impact parameter distributions,
results are presented in terms of
$b_{\mrm{MPI}}^{\mrm{norm}} = b / b_{\mrm{avg}}$, where
\begin{equation}
  b_{\mrm{avg}} =
  \frac{\int b \, \pint \, \d^2b}
       {\int \pint \, \d^2b} =
  \frac{1}{\snd}
  \int b \, \pint \, \d^2b
  ~,
\end{equation}
such that the average value is unity for minimum-bias events. Also of
interest is the enhancement factor associated with each interaction,
$\olap$ of eq.~(\ref{eq:nbb}). This is also normalised such
that the average is unity for the hard process in minimum-bias events,
$e_{\mrm{hard}}^{\mrm{norm}} = \olap / e_{\mrm{avg}}$, where
\begin{equation}
  e_{\mrm{avg}} =
  \frac{\int \nbb \, \pint \, \d^2b}{\int \nbb \, \d^2b} =
  \frac{1}{\sh} \int \nbb \, \pint \, \d^2b ~,
\end{equation}
compensating for the fact that the average number of interactions is raised
by removing the sample with no interactions.

As discussed in Sec.~\ref{sec:bsel}, those processes with final states
which cannot be produced by the MPI framework will begin their $\pT$
evolution at the kinematical limit. Due to this, their impact parameter
profiles will be picked directly according to $\olap \, \d^2b$, with no Sudakov
weighting. This offers a direct way to examine the effects of the
new matter profile in comparison to previous ones.

\subsubsection{$\Zz$ production}
\label{sec:Z0}
In this section, the MPI accompanying $\Zz$ (with no $\gamma^*$
interference) production is studied. The following matter profiles are
compared:\\
\renewcommand{\arraystretch}{1.15}
\indent\begin{tabular}{rl}
\textbf{SG:}      & single Gaussian, \\
\textbf{DG:}      & double Gaussian with default parameters $a_2 / a_1 = 0.4$
                    and $\beta = 0.5$,\\
\textbf{Overlap:} & overlap function with $p = 1.5$, \\
\textbf{Log:}     & logarithmically $x$-dependent Gaussian with
                    $a_1 = 0.15$.
\end{tabular}\\
This process is studied in $\p\p$ collisions at $\sqrt{s} = 7\TeV$. To
study only the effects of the MPI model, parton showers are switched
off. This does affect some longitudinal correlations; initial-state
radiation, in particular, is in competition with MPI for momentum from the
beams.

\begin{figure}
\centering
\includegraphics[scale=0.63,angle=270]{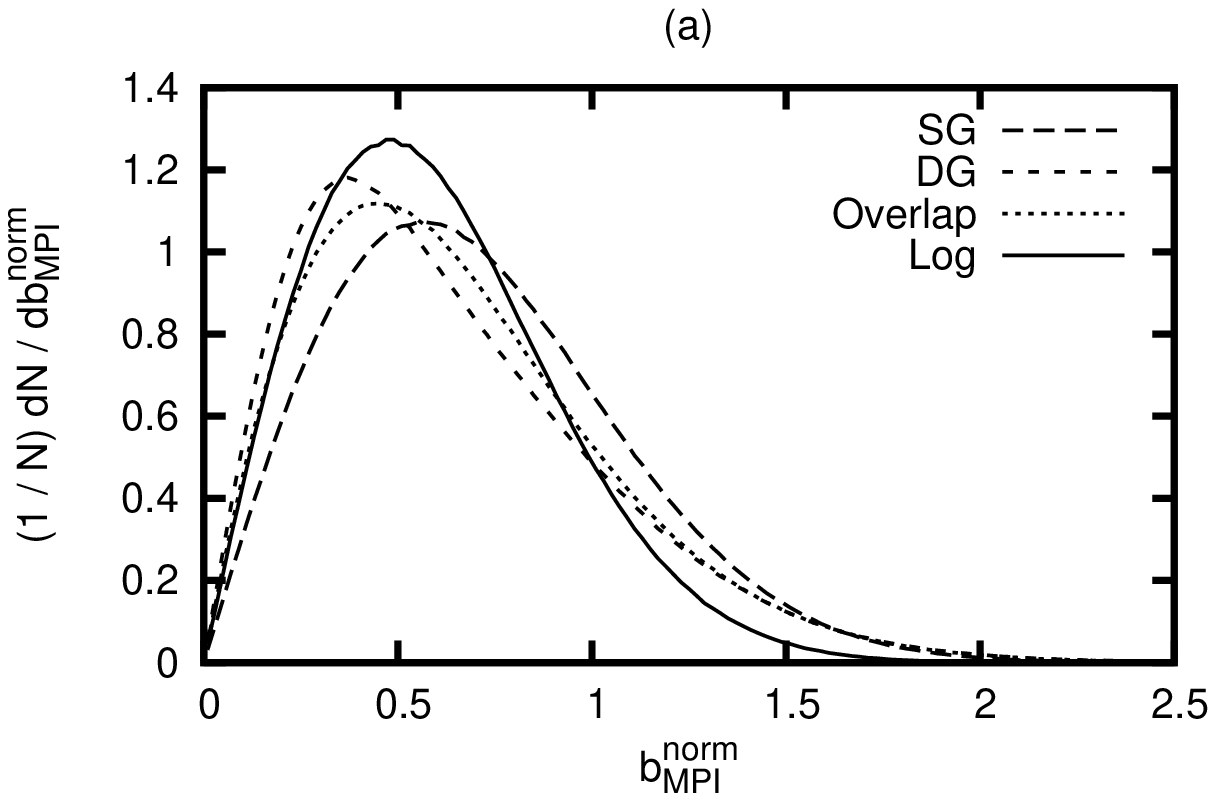}
\includegraphics[scale=0.63,angle=270]{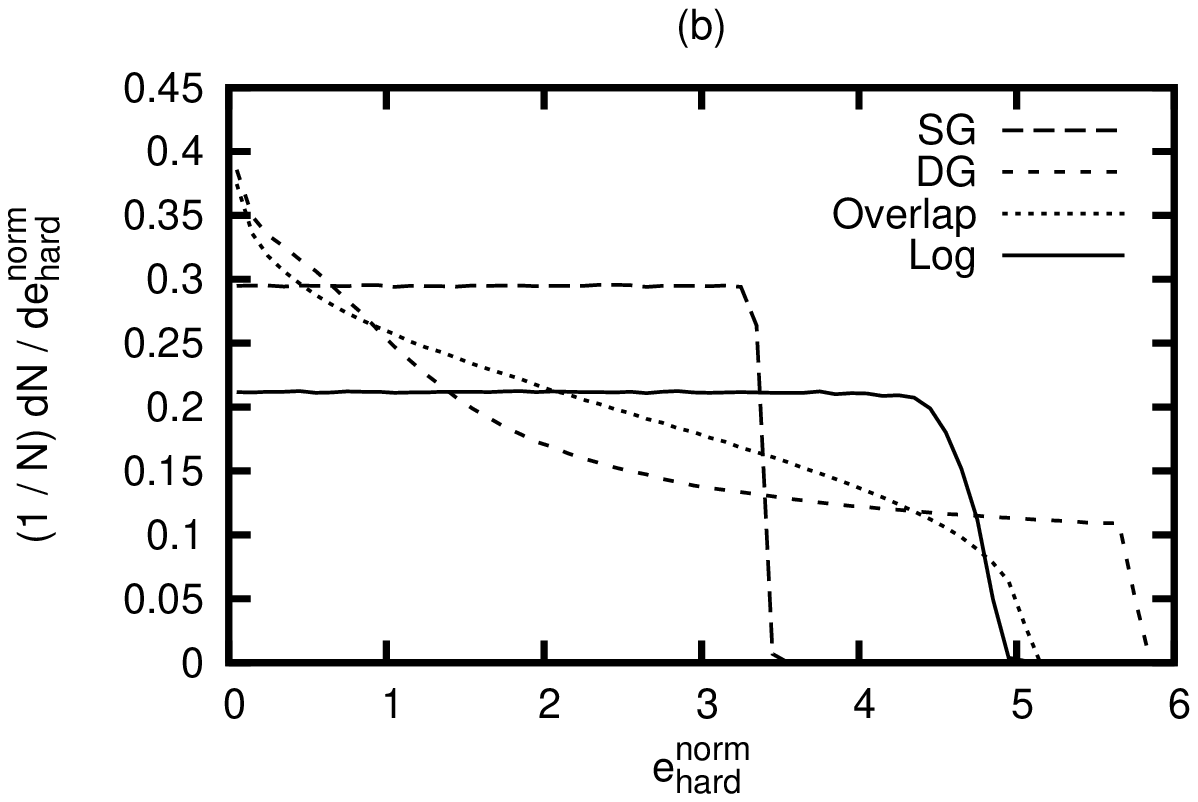}
\includegraphics[scale=0.63,angle=270]{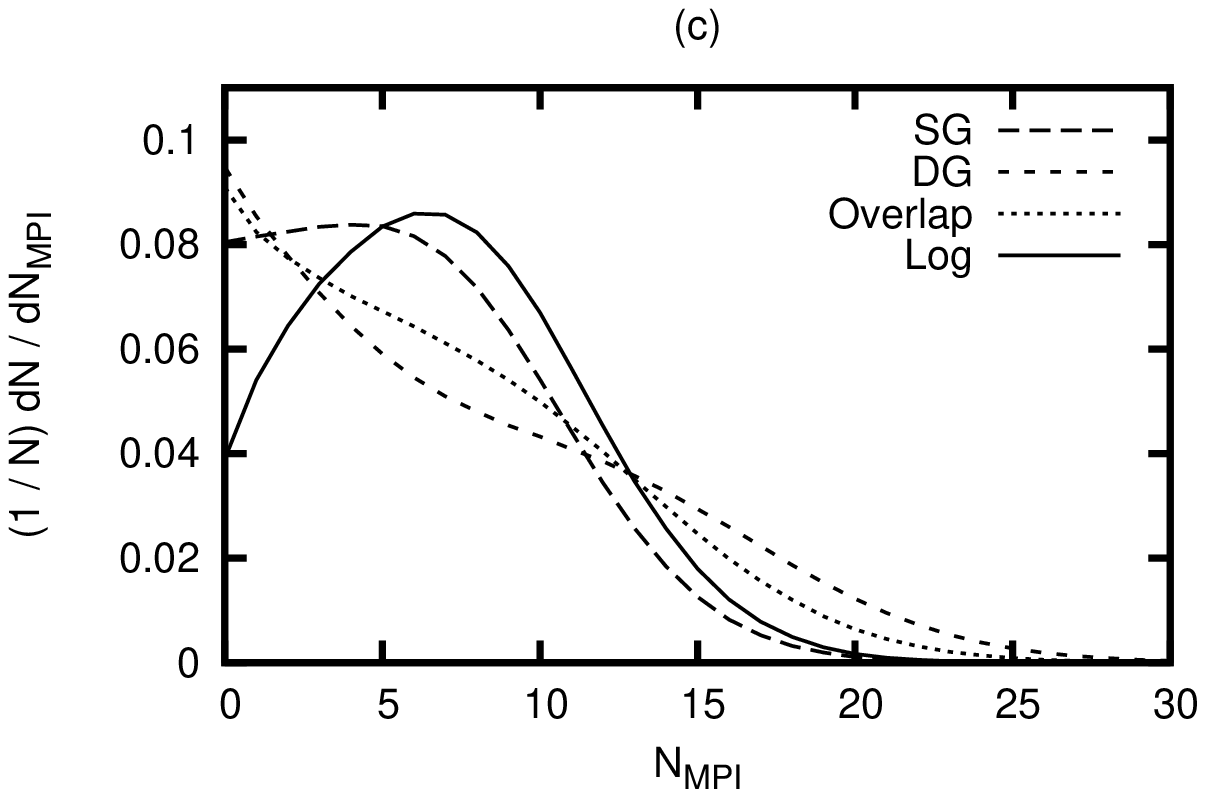}
\includegraphics[scale=0.63,angle=270]{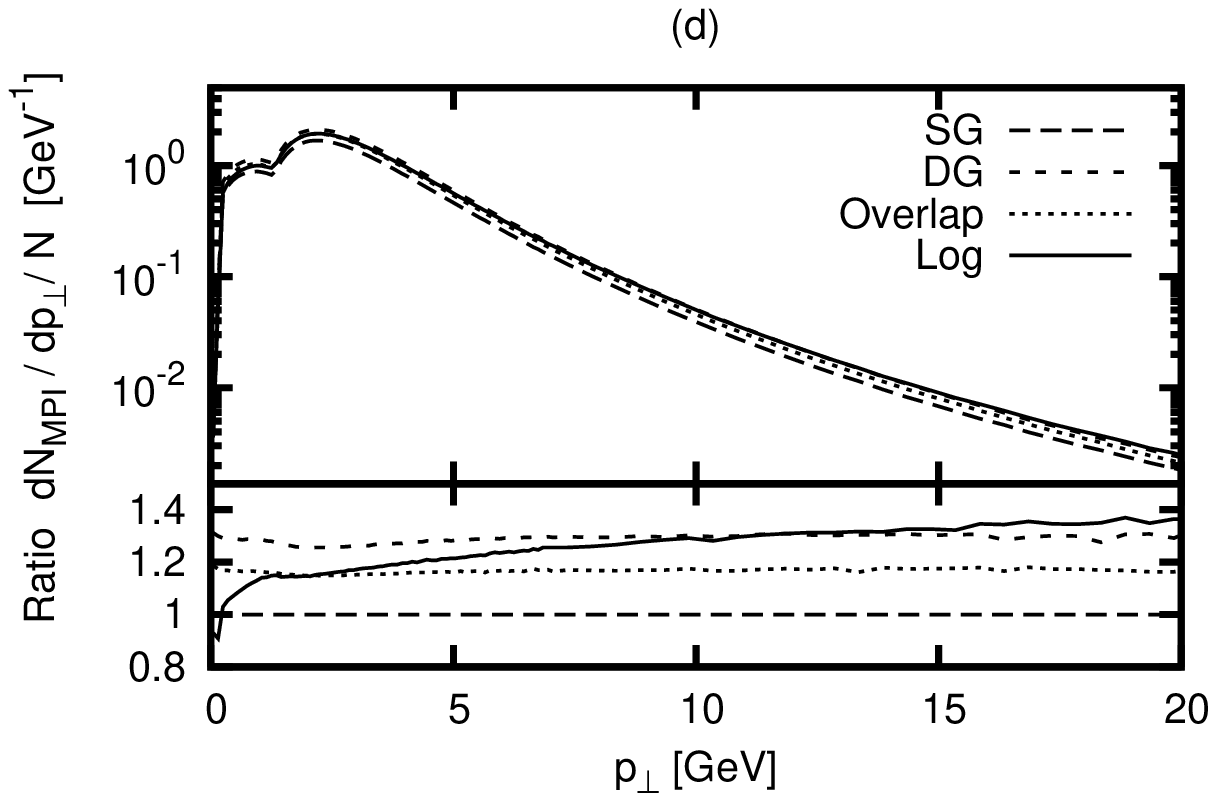}
\caption{$\Zz$ production in $\p\p$ collisions at $7\TeV$. (a) The impact
parameter distribution, (b) enhancement factor of the hard interaction, (c)
number of MPI and (d) inclusive $\pT$ spectrum of MPI per event.
The ratio plot in (d) is normalised to the single Gaussian result
\label{fig:Z0}}
\end{figure}

Fig.~\ref{fig:Z0}a shows the impact parameter distributions for these matter
profiles. As outlined previously, in $\Zz$ production, they are picked
unmodified from $\olap \, \d^2b$, but shown normalised such that the
average value would be unity for minimum bias events. Noting that
$\d^2b \propto b \, \d b$, it is possible to study the general features of
$\olap$ itself. In the double Gaussian profile, relative to the single
Gaussian, the central core of hadronic matter dominates at small $b$,
giving larger overlap values, although with a faster fall-off. The
peripheral Gaussian component then slows this fall-off, giving
contributions out to larger $b$ values. The overlap scenario sits roughly
between the single and double Gaussian distributions. The log profile is
now an average of single Gaussian overlap functions, whose widths are
determined by the combination of $x_1$ and $x_2$ values that contribute.
These $x$ values give a narrower distribution than the single Gaussian
case. At larger values of $a_1$, this distribution would become even more
narrow. It is these features which directly give rise to the form of the
impact parameter profile, shown in the figure. In particular, the log
scenario is peaked at smaller values than the single Gaussian, but without
a large tail out to high $b$ values, unlike the double Gaussian.

In Fig.~\ref{fig:Z0}b the distribution of the enhancement factor, $\olap$,
is shown, again normalised such the average value would be unity for
minimum bias events. It is noted that this distribution gives the overlap
for those events where $b$ has already been selected according to $\olap \,
\d^2 b$. Formally, this is stated by
$\d N / \d \olap = \d N / \d b * \d b / \d \olap$, where
$\d N / \d b \propto b \, \olap$. For the single Gaussian, where also
$\d \olap / \d b \propto b \, \olap$, the enhancement is flat. It stretches
from 0 to $\pi$, as determined by the normalisation of eq.~(\ref{eq:olap}).
The log profile is again an average of single Gaussians. The upper cutoff
is determined by the combination of $x$ values that gives the lowest
possible $a^2(x_1) + a^2(x_2)$. The width and shape of the fall-off is
related to the range and relative rate of these $x$ combinations,
respectively.  The shape for the overlap scenario can similarly be
calculated. For an overlap $\ol = k \, \exp(-b^{1.5})$, this is given by
$\d N / \d \ol \propto -\ln^{1/3}(k / \ol)$. It is not so easily calculated
for the double Gaussian, but the shape of the distribution can be
understood as its peripheral Gaussian component, stretching out to large
$b$, giving the peaking behaviour as $e_{\mrm{hard}}^{\mrm{norm}} \to 0$.

It is perhaps the number of MPI accompanying $\Zz$ production and their
$\pT$ spectrum that are of more interest, since they directly
influence physical observables. The distribution of the number of MPI
per event is shown in Fig.~\ref{fig:Z0}c. For the single Gaussian, double
Gaussian and overlap scenarios, the enhancement factor of the hard process
is retained for the remaining sequence of MPI. Before taking into account
energy-momentum conservation, the average number of MPI per event is
directly proportional to the enhancement factor, with the actual
number fluctuating around this mean value. The PDF rescaling then suppresses
the high tail of this distribution, pushing events to smaller
$N_{\mrm{MPI}}$. This does affect the overall shape of the curves,
but it remains true that the widths of the $N_{\mrm{MPI}}$ distributions
are essentially dictated by the widths of $e_{\mrm{hard}}^{\mrm{norm}}$.
For the double Gaussian and overlap scenarios, the peaking behaviour as
$e_{\mrm{hard}}^{\mrm{norm}} \to 0$ gives a similarly peaked distribution
as $N_{\mrm{MPI}} \to 0$.

For the log profile, the narrower impact parameter gives rise to fewer
events with small numbers of MPI. The tail, however, does not go out much
beyond the single Gaussian, as the double Gaussian and overlap profiles do.
This can be explained by the fact that the enhancement factor for the
sequence of MPI is no longer fixed to the hard enhancement factor, but
varies as a function of the $x$ values of each individual interaction.

The dominant process in MPI is $t$-channel gluon exchange. The parton-level
process has no suppression at large-$x$ values, but is affected by PDF
factors. Given the requirement that $\tau = x_1 x_2 < 4 \pTs / s$, as
$\pT$ falls, the minimum $\tau$ also falls, opening up new regions of
allowed $x$ values. From the small-$x$ peaking of the PDFs, one would expect
that, on average, the $x$ values will fall as $\pT$ does. This, in
turn, would lead to lower enhancement factors, as the partons become more
smeared out in the proton.

The result of the above is visible in Fig.~\ref{fig:Z0}d, where the
increase in the inclusive $\pT$ spectra of MPI is flat for the double
Gaussian and overlap profiles, relative to the single Gaussian, but falls
off towards low $\pT$ for the log scenario. The slope of this fall-off is
affected by the impact parameter distribution, which, in turn, is
determined by the $x$ values of the hard process. There are also additional
PDF rescaling effects at play, but checks show that these are small. The
change in shape of the absolute distributions at $\sim 2\GeV$ is due to the
freezing of the PDFs. Overall, then, the changing enhancement factor of MPI
regulates the amount of MPI, affecting the tail of the log profile in
Fig.~\ref{fig:Z0}c.

\subsection{Low mass Drell-Yan, $\Zz$ and $\Znz'$ production}

\begin{figure}
\centering
\includegraphics[scale=0.63,angle=270]{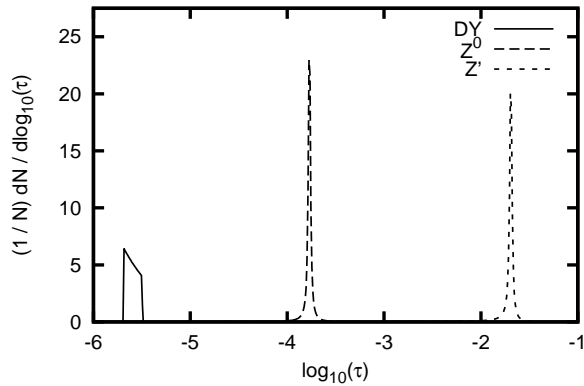}
\caption{The $\tau$ distribution of low mass Drell-Yan (DY), $\Zz$ and
$\Znz'$ events in $\p\p$ collisions at $\sqrt{s} = 7\TeV$
\label{fig:tau}}
\end{figure}

The next step of our comparisons is to include other processes, again where
there is no Sudakov involved, but for which the log profile will
dynamically produce variations in the distributions. In particular, the
following processes are used as templates to explore different
well-defined $x$ ranges:\\
\renewcommand{\arraystretch}{1.15}
\indent\begin{tabular}{rl}
\phantom{AAA}\textbf{DY:} & low mass Drell-Yan ($10.0 < \hat{m} < 12.5 \GeV$), \\
$\bf Z^0$\textbf{:}       & as in the previous section, \\
$\bf Z'$\textbf{:}        & a $1\TeV$ $\Znz'$ resonance.
\end{tabular}\\
Again, we note that parton showers are not switched on here, which also
take momentum from the beams, affecting PDF rescaling.
For the log profile, it is the combination of $x_1$ and $x_2$
together in eq.~(\ref{eq:loggauss}) that is important. In
Fig.~\ref{fig:tau}, the $\tau = x_1 x_2$ distribution of the three
processes is shown, with each contained in a well defined region.

In Sec.~\ref{sec:bsel}, some discussion was given relating to the choice of
impact parameter in hard processes, when its scale is in regions where the
Sudakov of eq.~(\ref{eq:softpT}) begins to vary away from unity, and where
the chance of having a harder MPI also grows. The Drell-Yan
process used here will be affected by these issues. We side step them here;
this comparison is designed to highlight the effects of the log profile in
certain $x$-ranges, without the additional complications of the Sudakov
factor. We retain the decision to pick impact parameters according to the
hard process and without any Sudakov weighting.

\begin{figure}
\centering
\includegraphics[scale=0.63,angle=270]{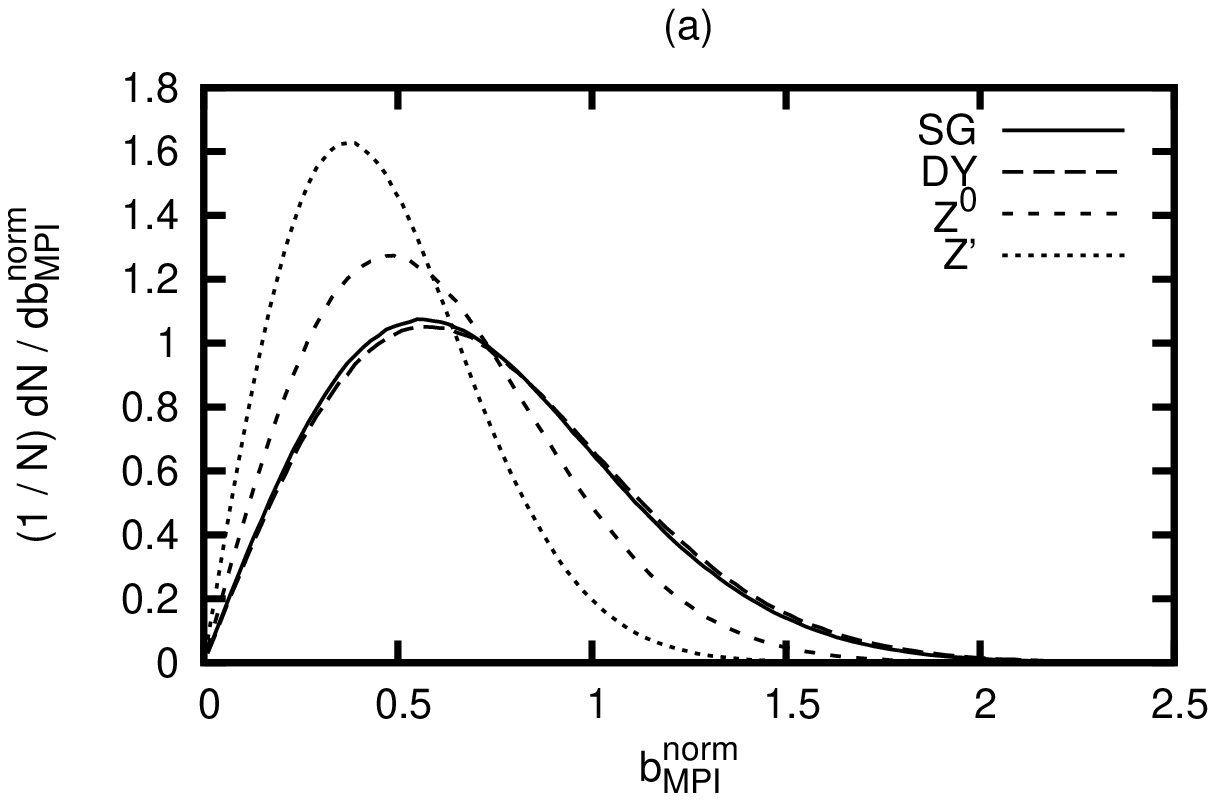}
\includegraphics[scale=0.63,angle=270]{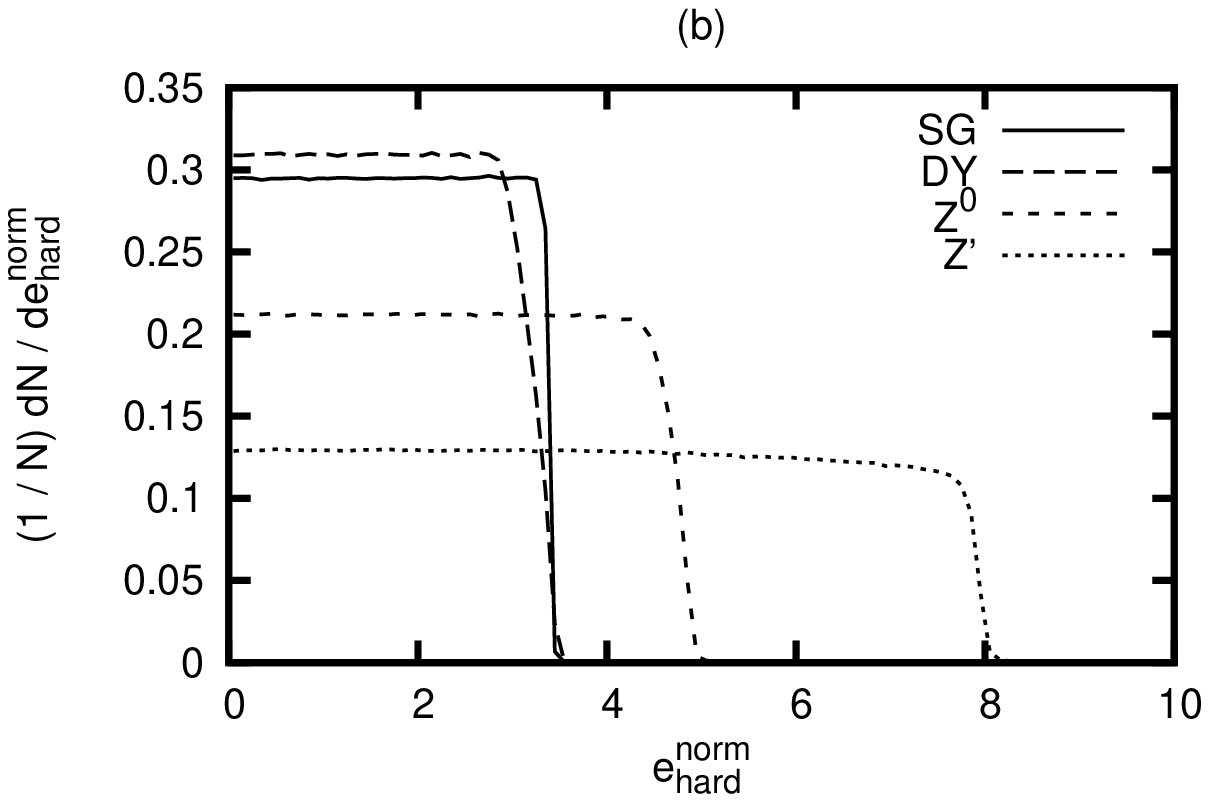}
\caption{Low-mass Drell-Yan, $\Zz$ and $\Znz'$ production in $\p\p$
collisions at $\sqrt{s} = 7\TeV$. (a) The impact parameter distribution and
(b) enhancement factor of the hard interaction per event. The single
Gaussian distributions are identical between the three processes
\label{fig:DYZZ}}
\end{figure}

\begin{figure}
\centering
\includegraphics[scale=0.63,angle=270]{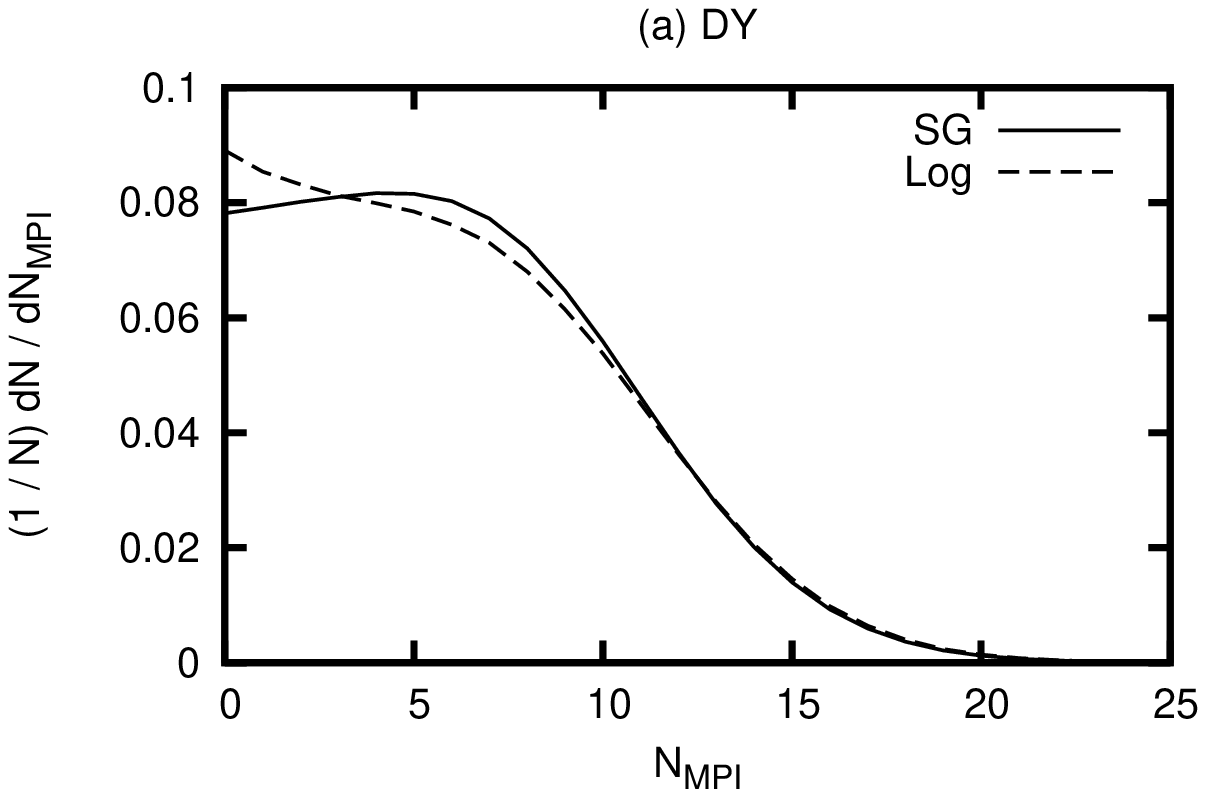}
\includegraphics[scale=0.63,angle=270]{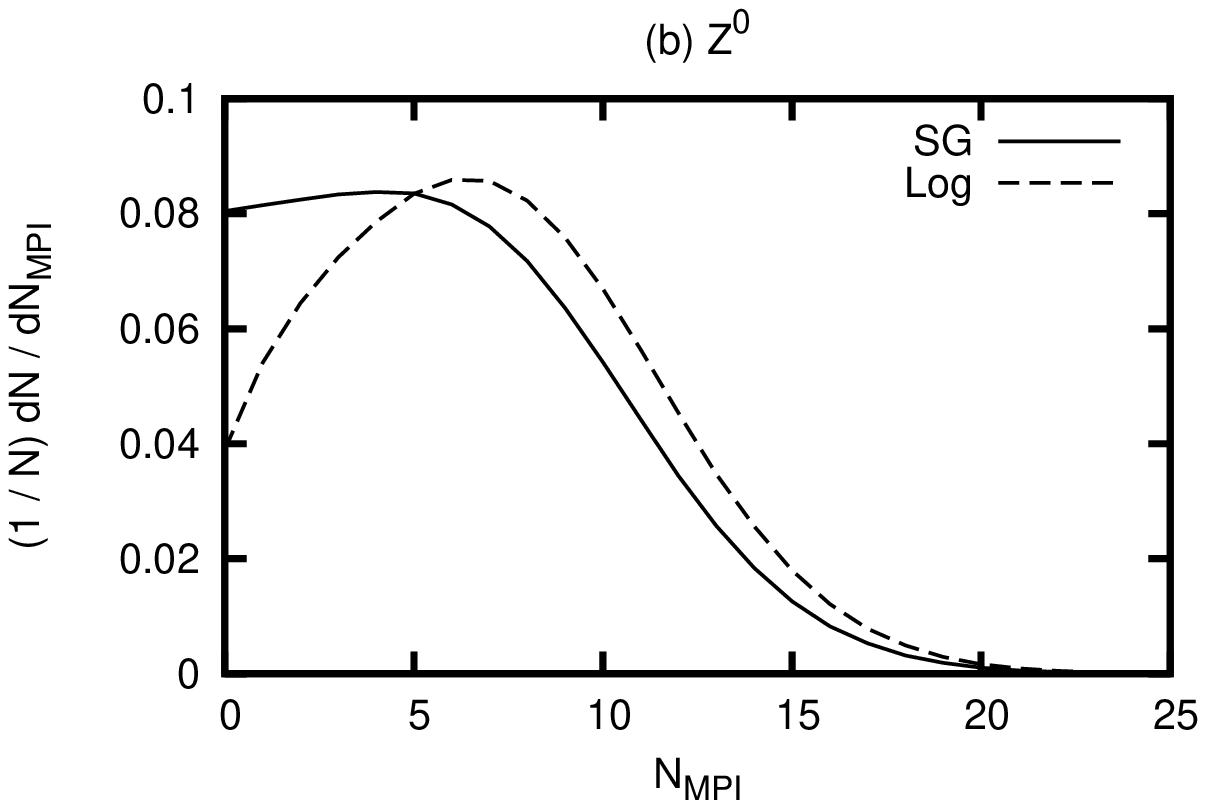}
\includegraphics[scale=0.63,angle=270]{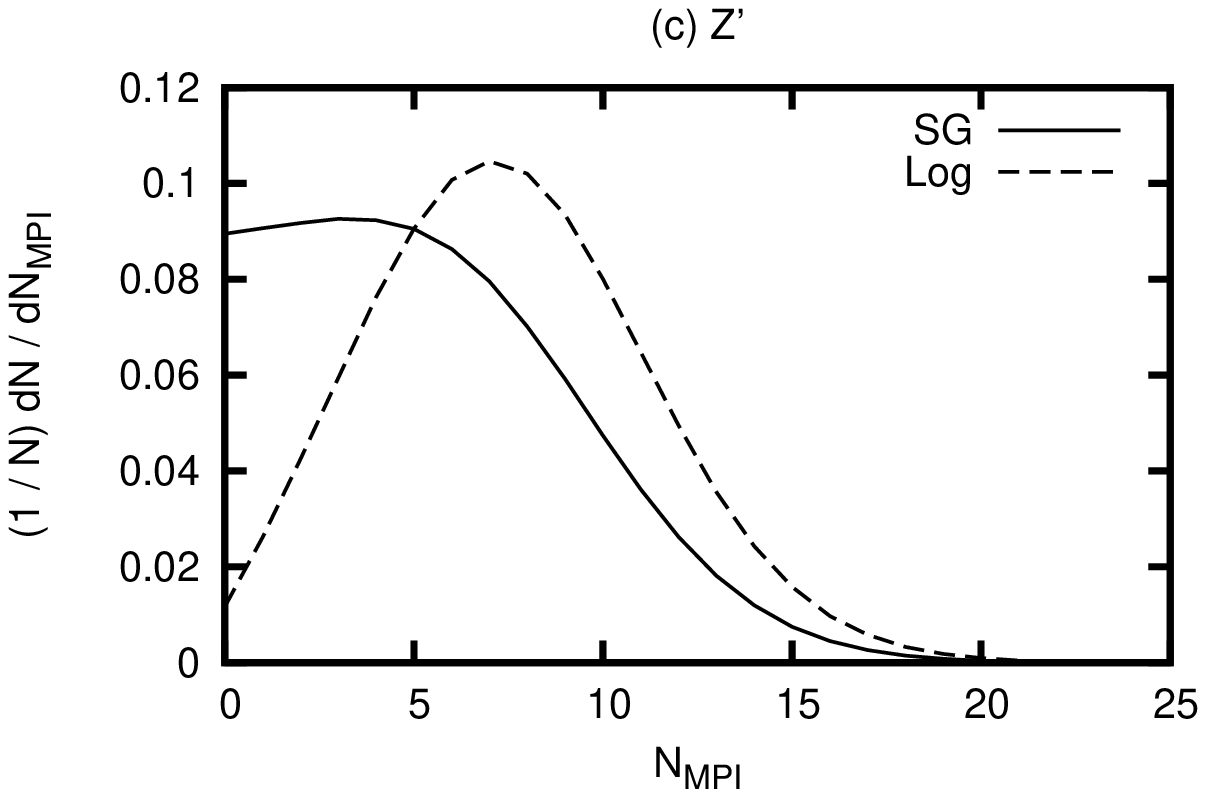}
\caption{The number of MPI per event in (a) low-mass Drell-Yan, (b) $\Zz$ and
(c) $\Znz'$ production
\label{fig:DYZZnMI}}
\end{figure}

\begin{figure}
\centering
\includegraphics[scale=0.63,angle=270]{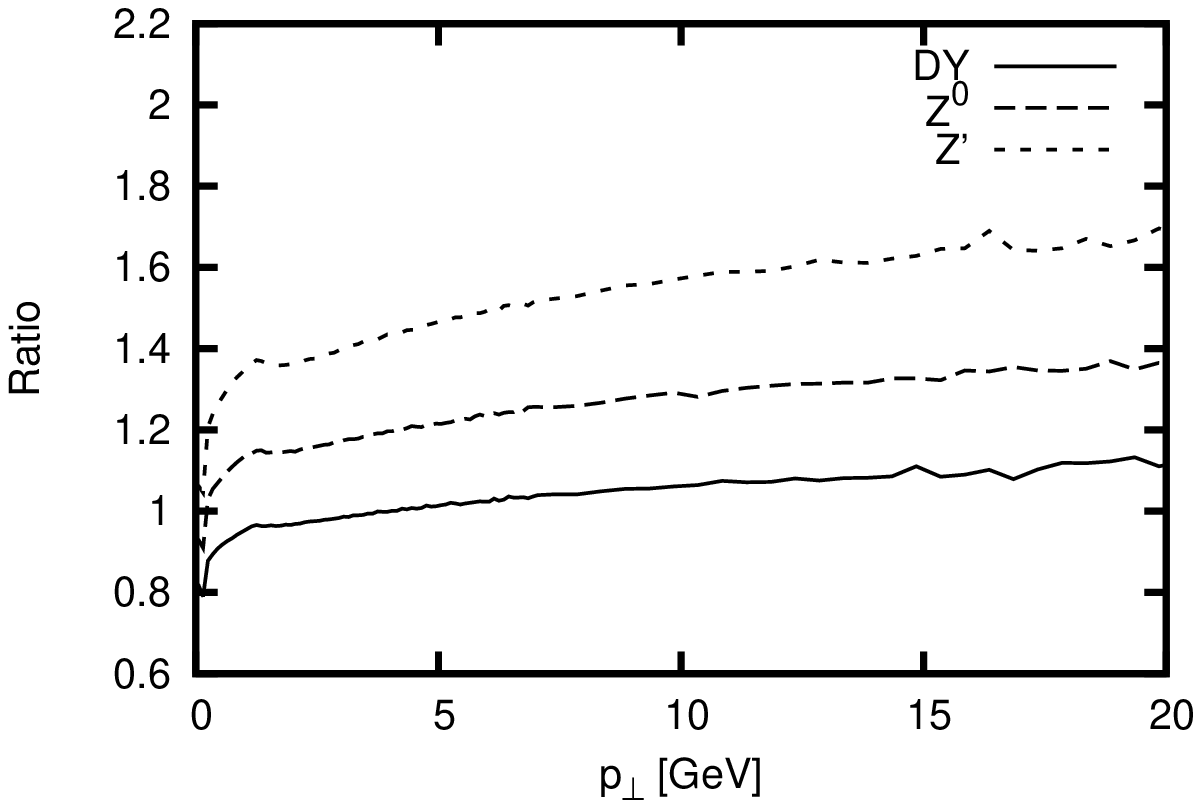}
\caption{Ratio of the inclusive $\pT$ spectrum of MPI for the log profile
to the single Gaussian result for the three different processes
\label{fig:DYZZpTMI}}
\end{figure}

Fig.~\ref{fig:DYZZ} shows the (a) impact parameter and (b) enhancement
factor in the hard process per event, this time comparing the processes given
above. These are also compared to the single Gaussian profile, which, as noted
previously, gives the same results for these distributions for all three
processes. In Fig.~\ref{fig:DYZZnMI}, the number of MPI
is shown for (a) low-mass Drell-Yan, (b) $\Zz$ and (c) $\Znz'$.
Fig.~\ref{fig:DYZZpTMI} shows the ratio of the inclusive $\pT$ spectrum of
MPI for the log profile to the single Gaussian result for the three
processes.


All the features in these distributions can be understood in terms of the
previous discussion of $\Zz$ events. As expected, the impact parameter
distributions become narrower as the $\tau$ range in question increases,
leading to a wider distribution for the enhancement factor in the hard
process. The Drell-Yan process gives very similar results to the single
Gaussian for the impact parameter, hard enhancement factor and number of
MPI distributions. The $x$ values that contribute, in some sense,
correspond to an average ``hardness'', with the same amount of activity
above and below. All three processes have a similar endpoint in
$N_\mrm{MPI}$, due to PDF rescaling. The even narrower impact parameter for
$\Znz'$ events further suppresses events with low $N_\mrm{MPI}$, relative
to $\Zz$ production.  The $\pT$ ratios of Fig.~\ref{fig:DYZZpTMI} all show
an enhancement of high-$\pT$ activity relative to low, as expected. As the
$x$ values of the hard process get larger, and the impact parameter profile
narrower, the slope becomes steeper.

\subsection{Minimum bias}

We now move on to minimum-bias events. Here, there are additional
correlations to consider, relative to the hard processes of the
previous section. In particular, once a hard process has been selected,
the subsequent MPI evolution will begin from this scale, meaning that
high-$\pT$ events are likely to have more MPI, given the larger $\pT$
evolution range they have available. Low-$\pT$ events will also be biased
towards larger impact parameters relative to high-$\pT$ ones, given the
Sudakov weighting of eq.~(\ref{eq:softpT}).

\begin{figure}
\centering
\includegraphics[scale=0.63,angle=270]{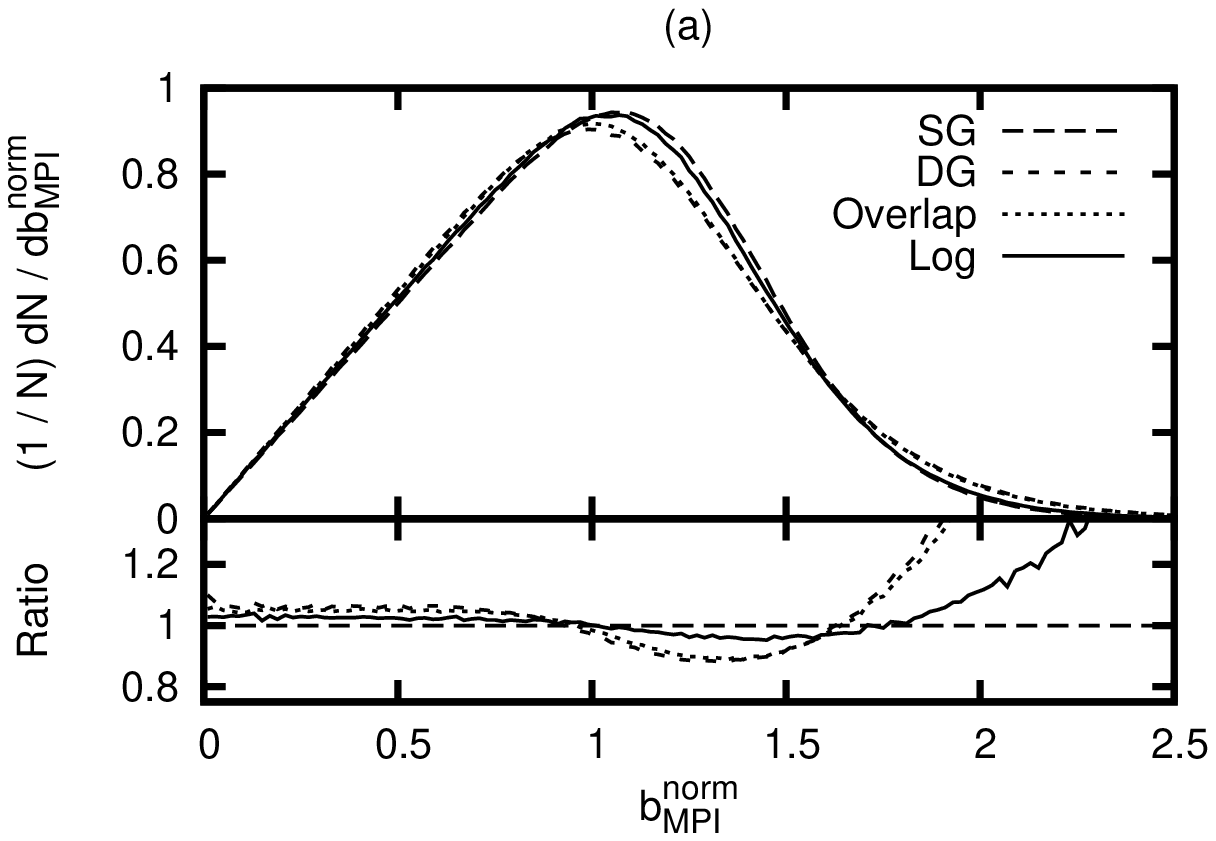}
\includegraphics[scale=0.63,angle=270]{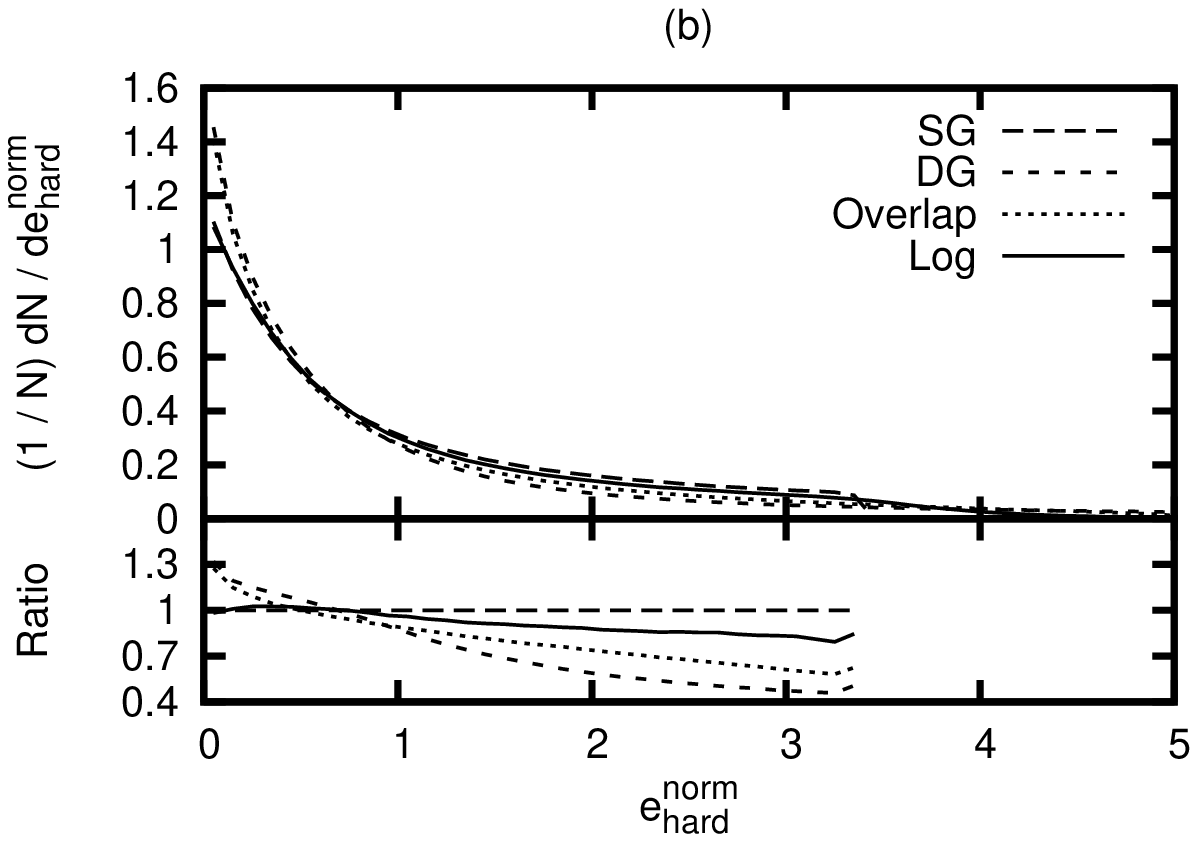}
\includegraphics[scale=0.63,angle=270]{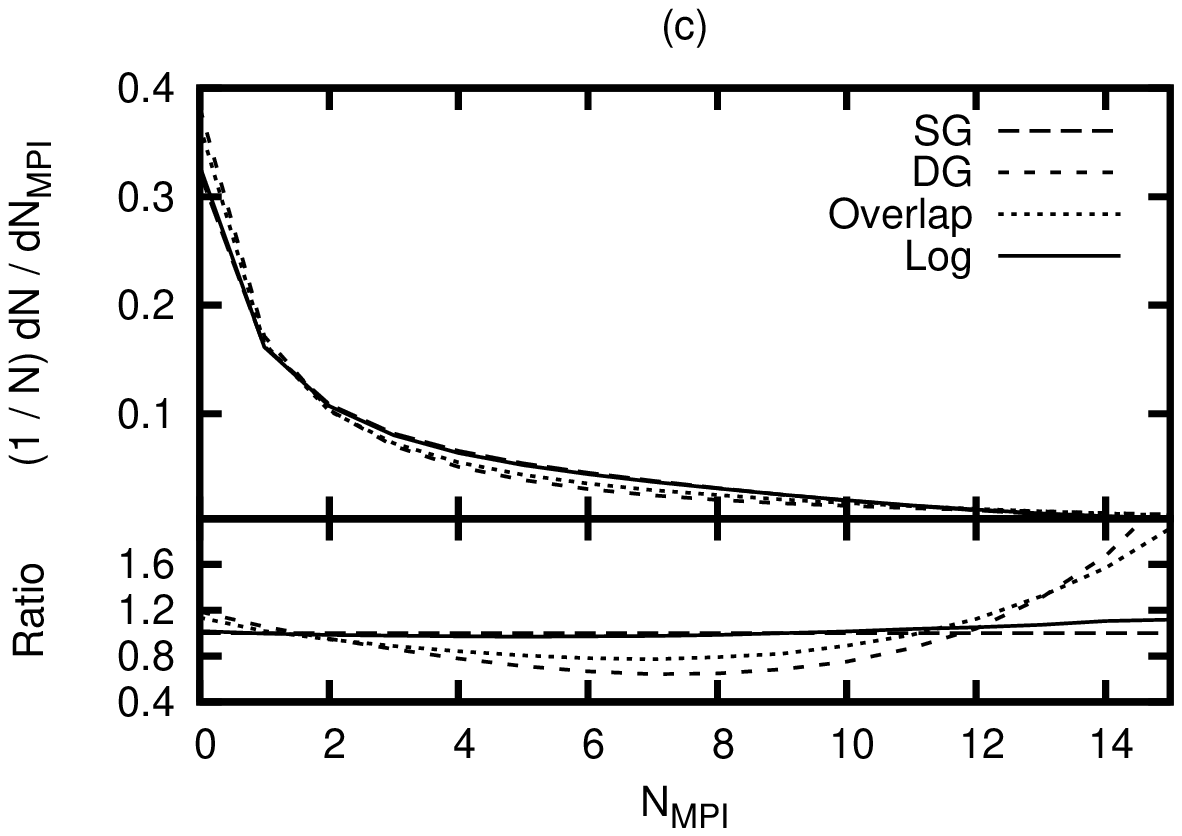}
\caption{Minimum-bias events in $\p\p$ collisions at $\sqrt{s} = 7\TeV$. 
(a) The impact parameter distribution, (b) enhancement factor of the hard
interaction and (c) number of MPI per event. Ratio plots are normalised to
the single Gaussian results
\label{fig:mbMI}}
\end{figure}

Again, we begin with the (a) impact parameter, (b) enhancement factor in
the hard process and (c) number of MPI per event, now shown in
Fig.~\ref{fig:mbMI}, for the same four matter profiles used in $\Zz$
production. The differences here are somewhat smaller than for the
hard processes, so the ratios of the double Gaussian, overlap and
log matter profiles to the single Gaussian are also given.

Both the impact parameter and enhancement factor distributions are now
fixed, so that they have an average value of unity for all the different
profiles. Any increase or decrease of these distributions over a given
range must be compensated elsewhere. The shape of the impact parameter
distributions are now directly dictated by $\nbb$. For the log profile, it
contains all the correlations brought about by the $x$-dependent width.
Relative to the single Gaussian, the other three profiles show the same
features; larger contributions at small and large $b$ values and a region
of intermediate $b$ where it sits below. For the log profile, this is
consistent with the shape of $\nbb$ described in Sec.~\ref{sec:psize}.
Fig.~\ref{fig:mbMI}a shows that the variation of the log
profile, relative to the single Gaussian, is smaller than for the double
Gaussian and overlap scenarios.

The overall shape of the $e_{\mrm{hard}}^{\mrm{norm}}$ distributions is
given by the effect of the impact parameter profiles, which now vary as a
function of the $\pT$ of the hard process, as noted above. Low-$\pT$
events, which dominate, will be biased to higher impact parameters, giving
the increase at low enhancement factors. Relative to the single Gaussian,
the three other profiles all show the same features; there is an increase
at low $e_{\mrm{hard}}^{\mrm{norm}}$, followed by a decrease, before the
tails continue out beyond where the single Gaussian cuts off. These changes
are a direct consequence of the changed impact parameter distributions.

For the $N_{\mrm{MPI}}$ distributions, the overall shape is now due to the
correlation between $\pT$ and the number of MPI. On average, the more
numerous low-$\pT$ events have less range of evolution, and therefore fewer
MPI. The variation of the log profile, relative to the single Gaussian is
small, but follows those of the double Gaussian and overlap; there is an
increase at low $N_{\mrm{MPI}}$, followed by a decrease, before again
increasing in the tails. Again, these changes follow directly from the
differences of the impact parameter distributions.

\begin{figure}
\centering
\includegraphics[scale=0.63,angle=270]{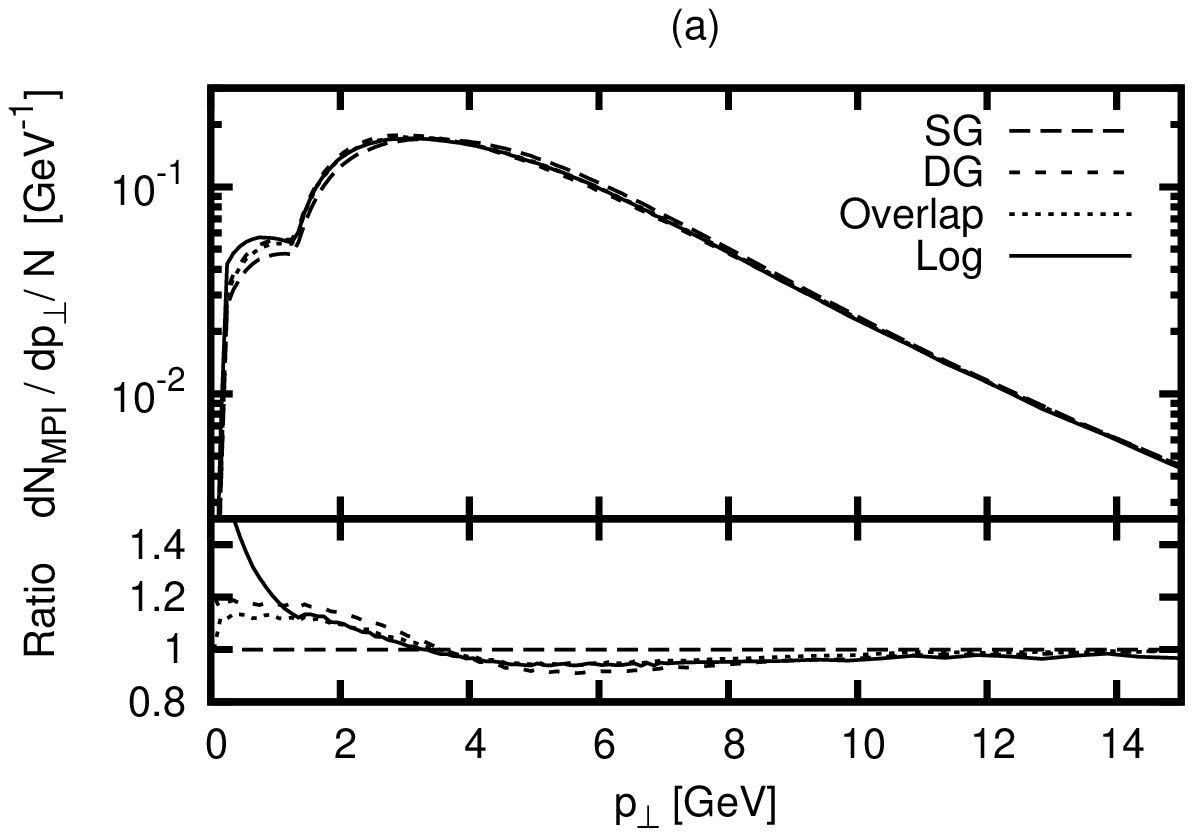}
\includegraphics[scale=0.63,angle=270]{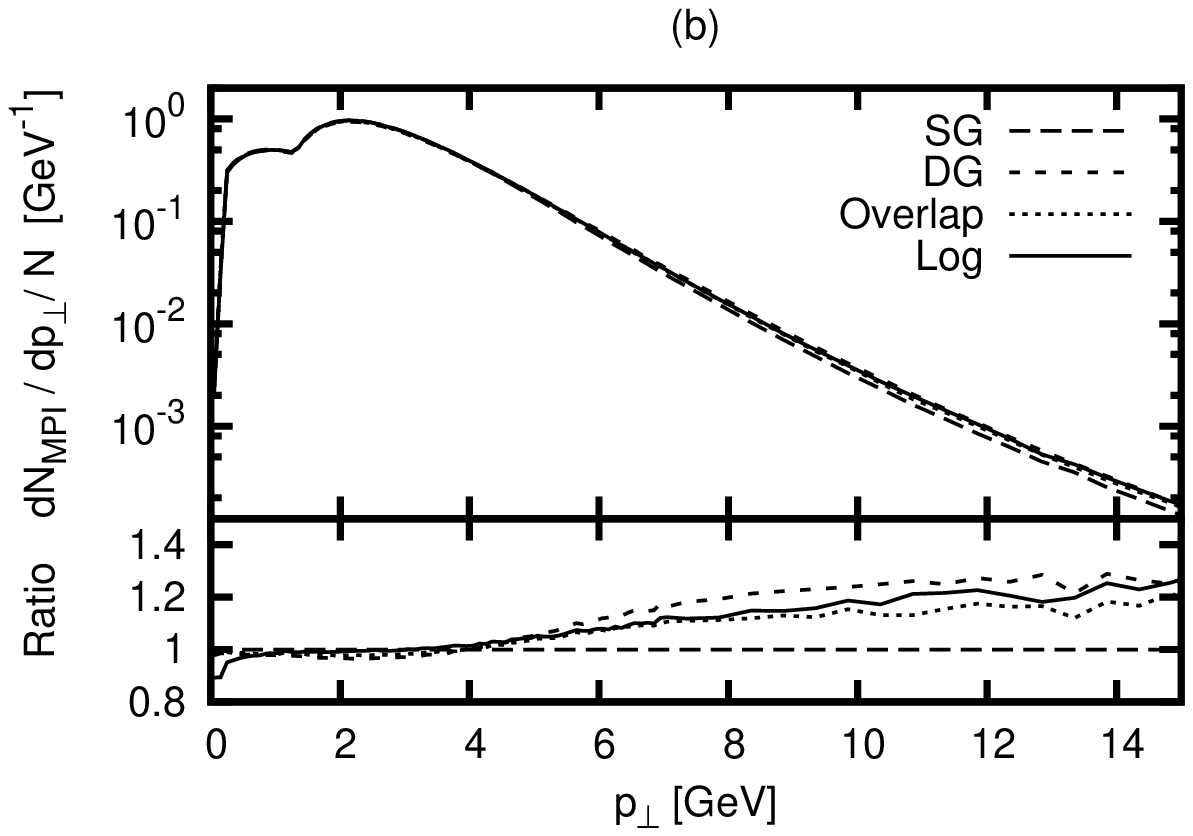}
\caption{$\pT$ distribution (a) of the hard process and (b) inclusive for
subsequent MPI in minimum bias events. Ratio plots are normalised to the
single Gaussian result
\label{fig:mbpT}}
\end{figure}

Finally, in Fig.~\ref{fig:mbpT}, the $\pT$ distribution (a) of the hard
process and (b) inclusive for subsequent MPI are shown.
For the hard process, relative to the single Gaussian, the other three
profiles give more activity at low $\pT$, relative to high. 
It becomes easier to have peripheral interactions involving small-$x$
partons, with the event containing no further activity. In the high-$\pT$
tails, the overlap is essentially saturated. There is a sharp
rise as $\pT \to 0$ for the log profile. This change in shape is
due to the freezing of the PDFs; in these low-$\pT$ bins, there is no
penalty to pay for taking higher $x$ values, up to this freezing point,
resulting in extra contributions here. The inclusive $\pT$ spectra for the
subsequent MPI now give exactly the opposite results to those of the hard
process, such that, when they are summed together, they give
back the unmodified $\pT$ spectrum of eq.~(\ref{eq:pTall}), as they must.

\subsection{Minimum-bias and underlying-event studies}

We can examine the effect of this new matter profile on
minimum-bias and underlying event studies. In particular, Tune 4C, used
also in the previous studies, offers an attractive starting point.
This tune is based on a modification to a Tevatron tune, such that it is
able to describe early LHC data. One of the features of this tune is a
single Gaussian matter profile, which gives a reasonable match to both the
rise of the underlying event as well as the width of charged multiplicity
distributions in minimum-bias events.

\begin{figure}
\centering
\includegraphics[scale=0.63,angle=270]{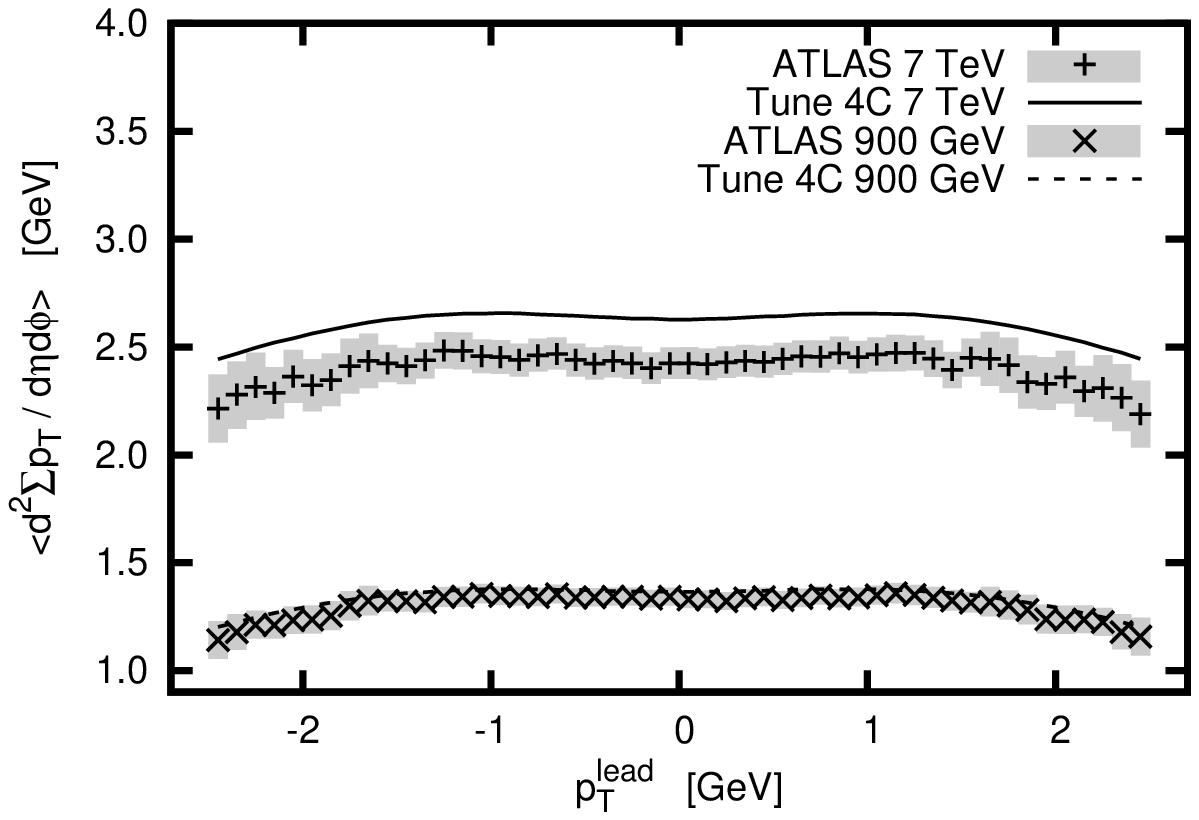}
\includegraphics[scale=0.63,angle=270]{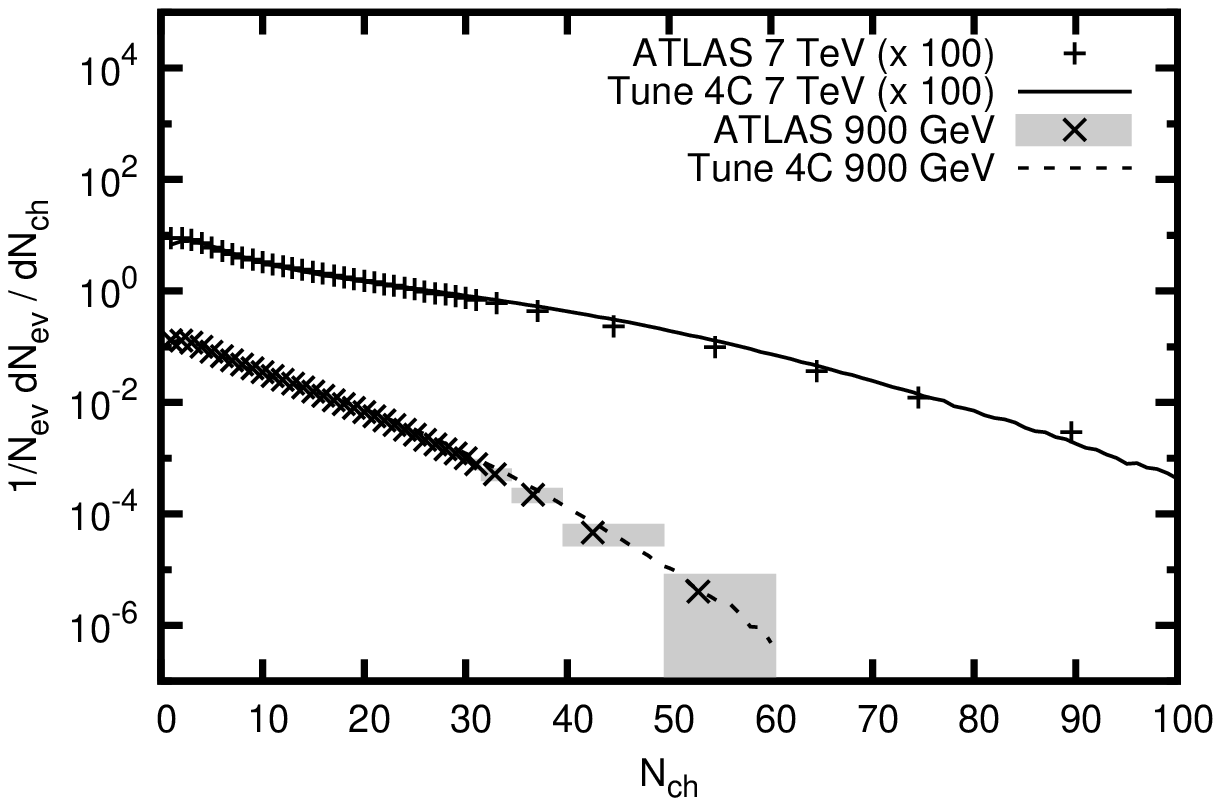}
\includegraphics[scale=0.63,angle=270]{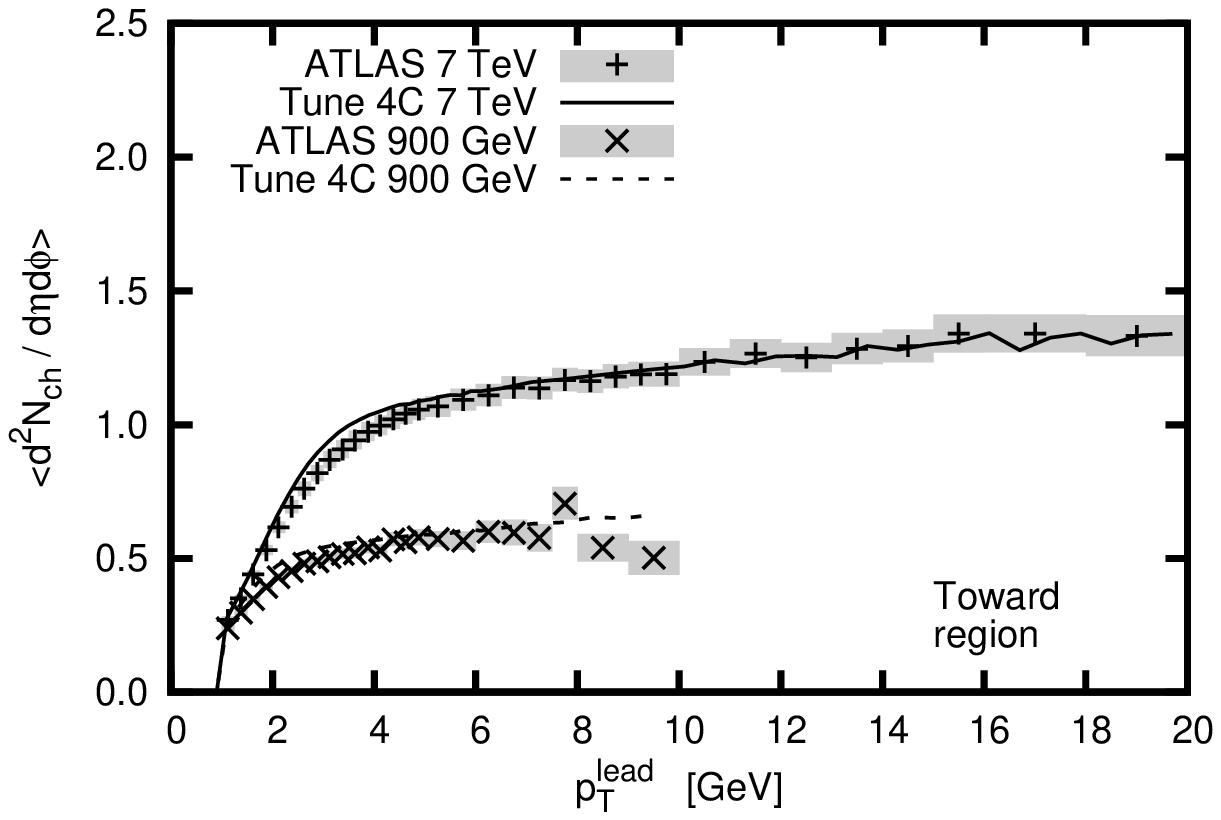}
\includegraphics[scale=0.63,angle=270]{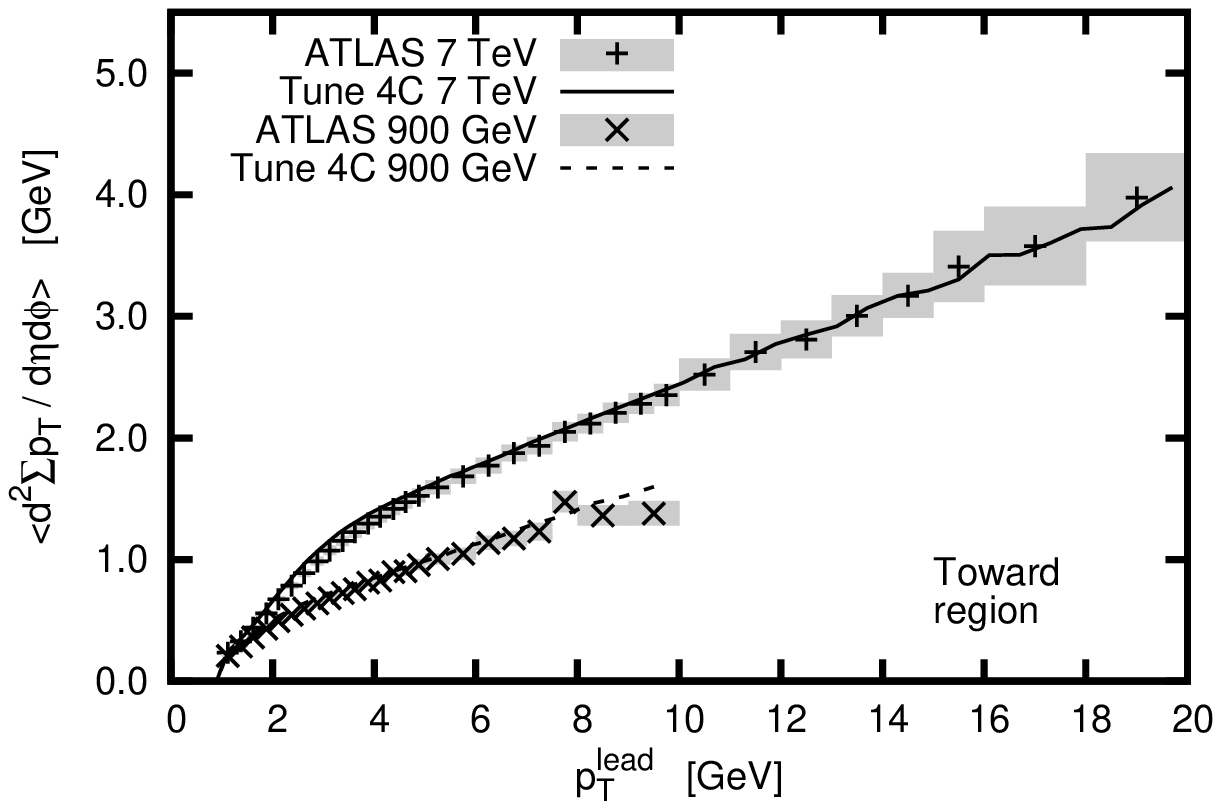}
\includegraphics[scale=0.63,angle=270]{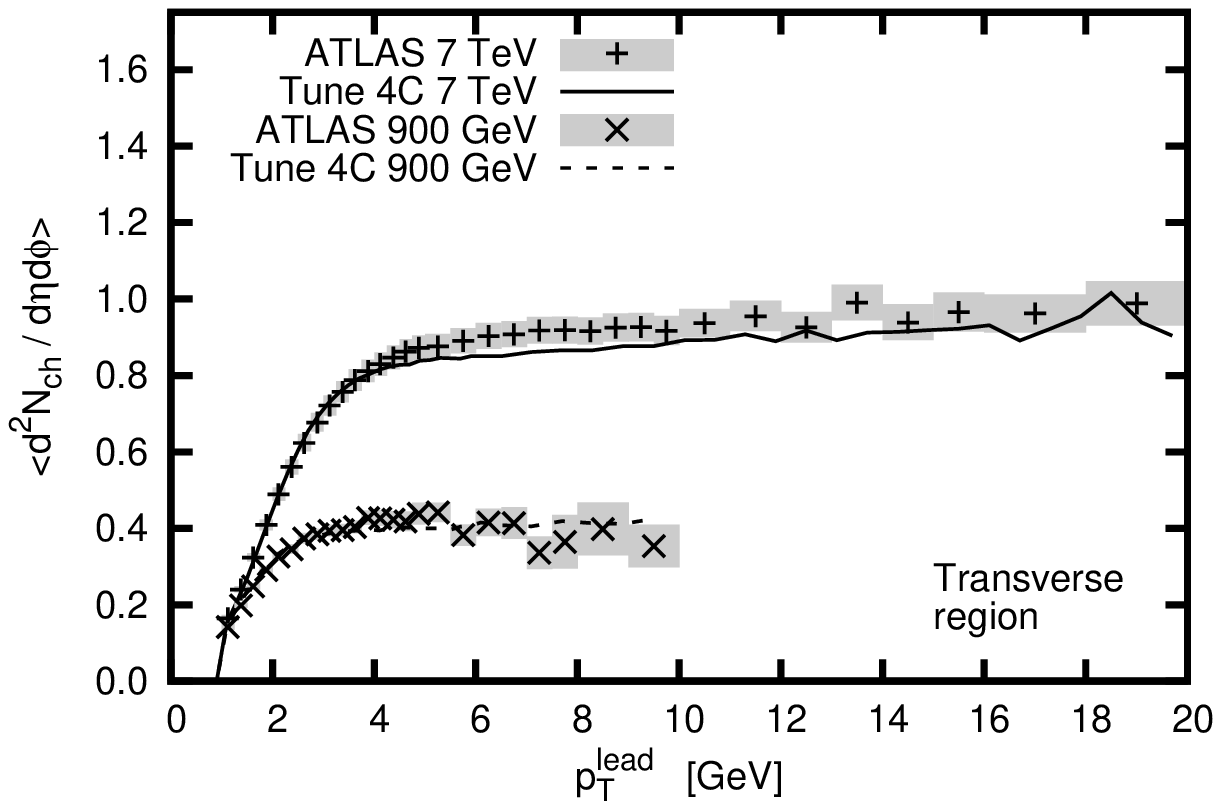}
\includegraphics[scale=0.63,angle=270]{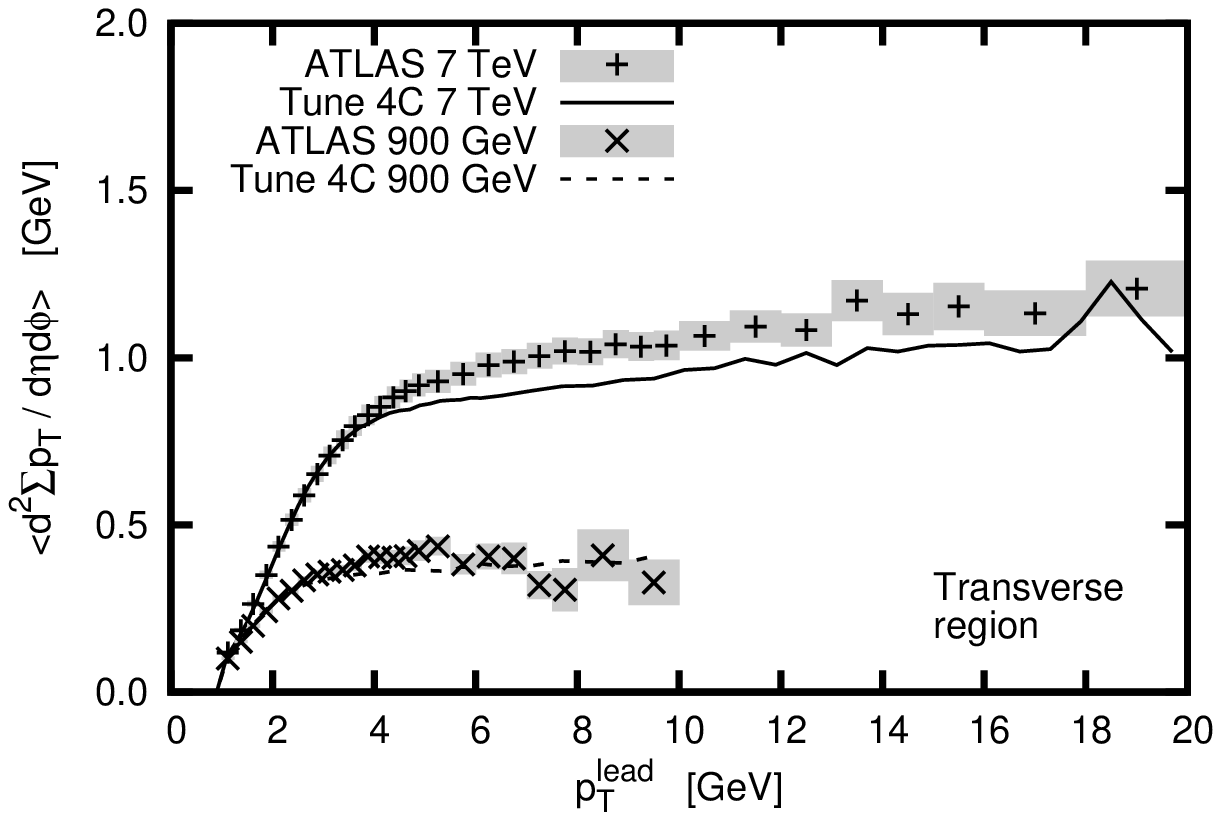}
\caption{Tune 4C compared against early LHC data. Further details are given
in the text
\label{fig:4C}}
\end{figure}

In Fig.~\ref{fig:4C}, the results of this tune are shown for
(left-to-right, top-to-bottom):
\begin{itemize}
\item[1)] ATLAS ($|\eta| < 2.5$, $\pT > 500\MeV$) INEL$>$0 minimum-bias
          dataset. Charged rapidity distribution in minimum-bias events
          at $\sqrt{s} = 900\GeV$ and $7\TeV$ \cite{Aad:2010rd,AtlasMB7}.
          The $900\GeV$ data is taken from the online HEPDATA database,
          while the $7\TeV$ data is taken from the corresponding
          reference.
\item[2)] As (1), but showing the charged multiplicity distributions.
          Errors are not included for the $7\TeV$ data.
\item[3)] ATLAS ($|\eta| < 2.5$, $\pT > 500\MeV$) charged track based
          underlying event at $\sqrt{s} = 900\GeV$ and
          $7\TeV$ \cite{AtlasUE}. A charged track of $\pT > 1\GeV$ in the
          $\eta$ acceptance is required to trigger an event. Data and
          errors have been read off from the corresponding reference.
          Charged particle number density in the toward region.
\item[4)] As (3), but showing the sum-$\pT$ density in the toward region.
\item[5)] As (3), but showing the charged particle number density in
          the transverse region.
\item[6)] As (3), but showing the sum-$\pT$ density in the transverse
          region.
\end{itemize}
Where errors are shown, they represent the systematic and statistical
errors summed in quadrature. Although the rise of the underlying event is
too steep in the toward region, and activity in the transverse region is
slightly too low, overall it gives a reasonable description of data.

\begin{figure}
\centering
\includegraphics[scale=0.63,angle=270]{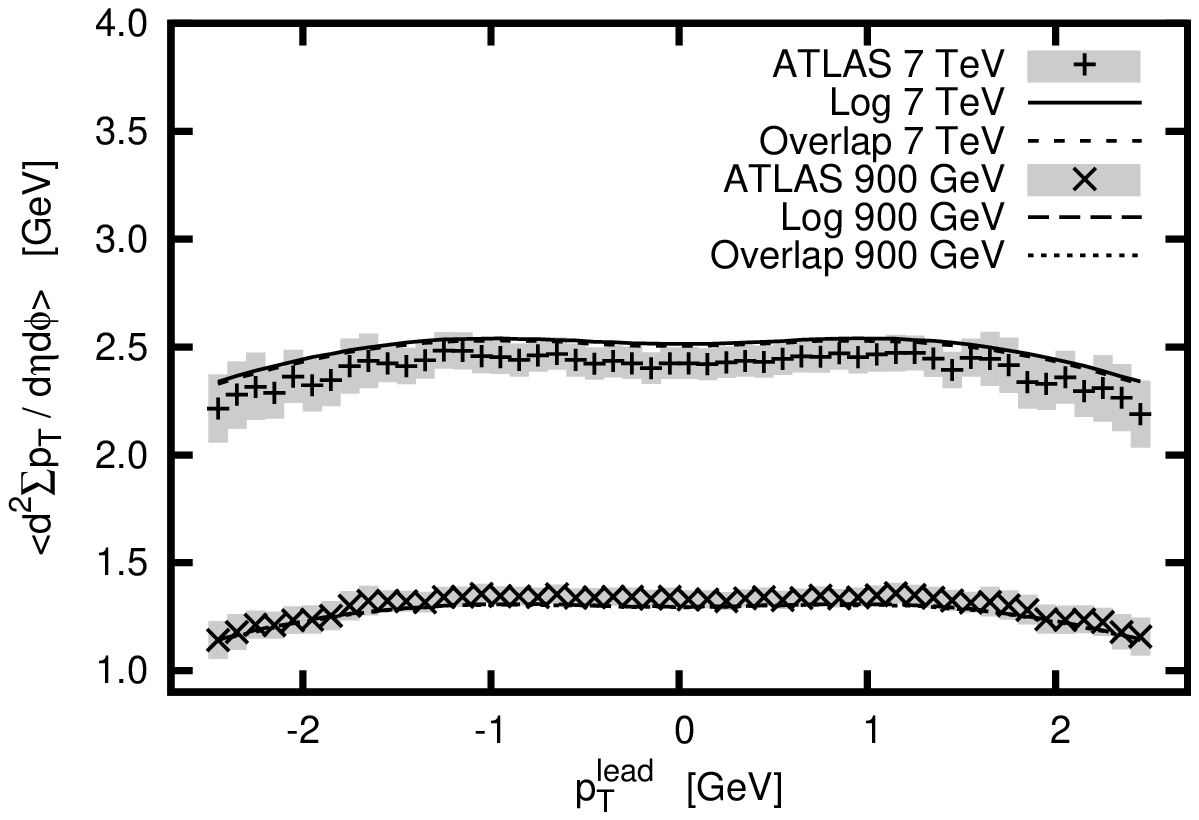}
\includegraphics[scale=0.63,angle=270]{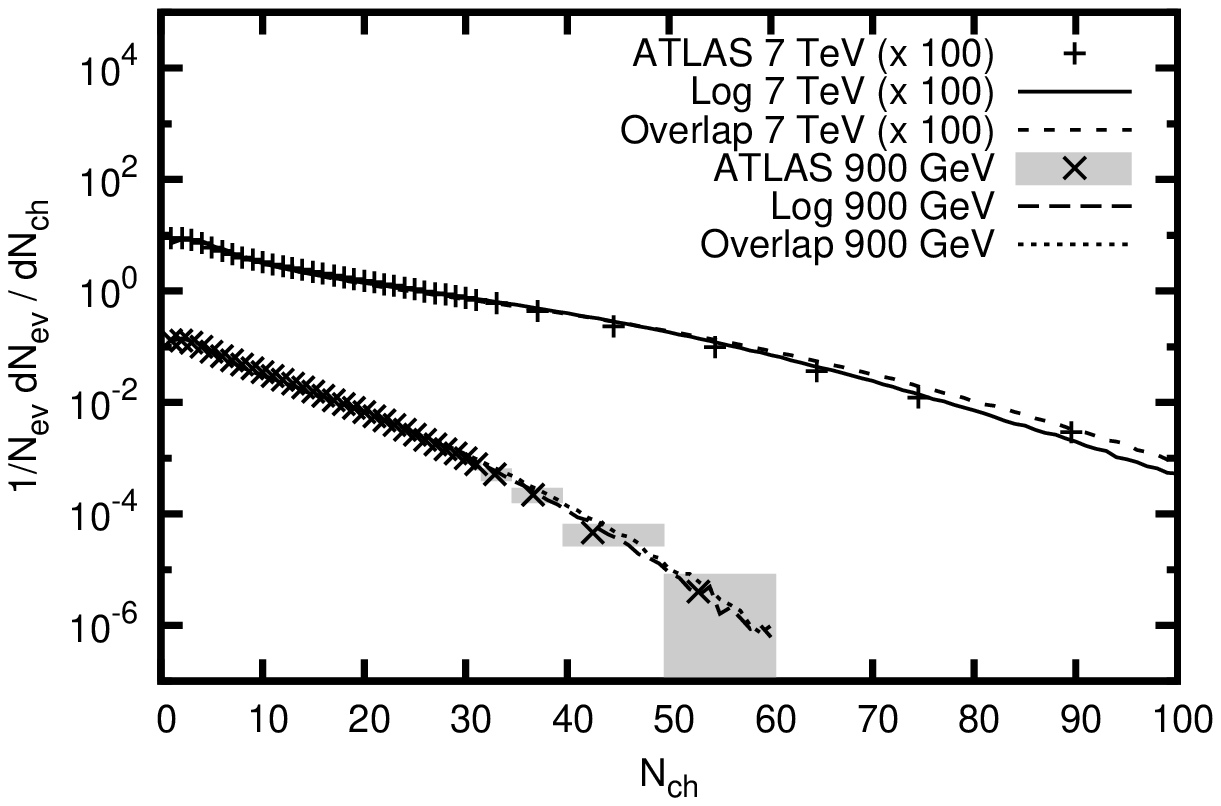}
\includegraphics[scale=0.63,angle=270]{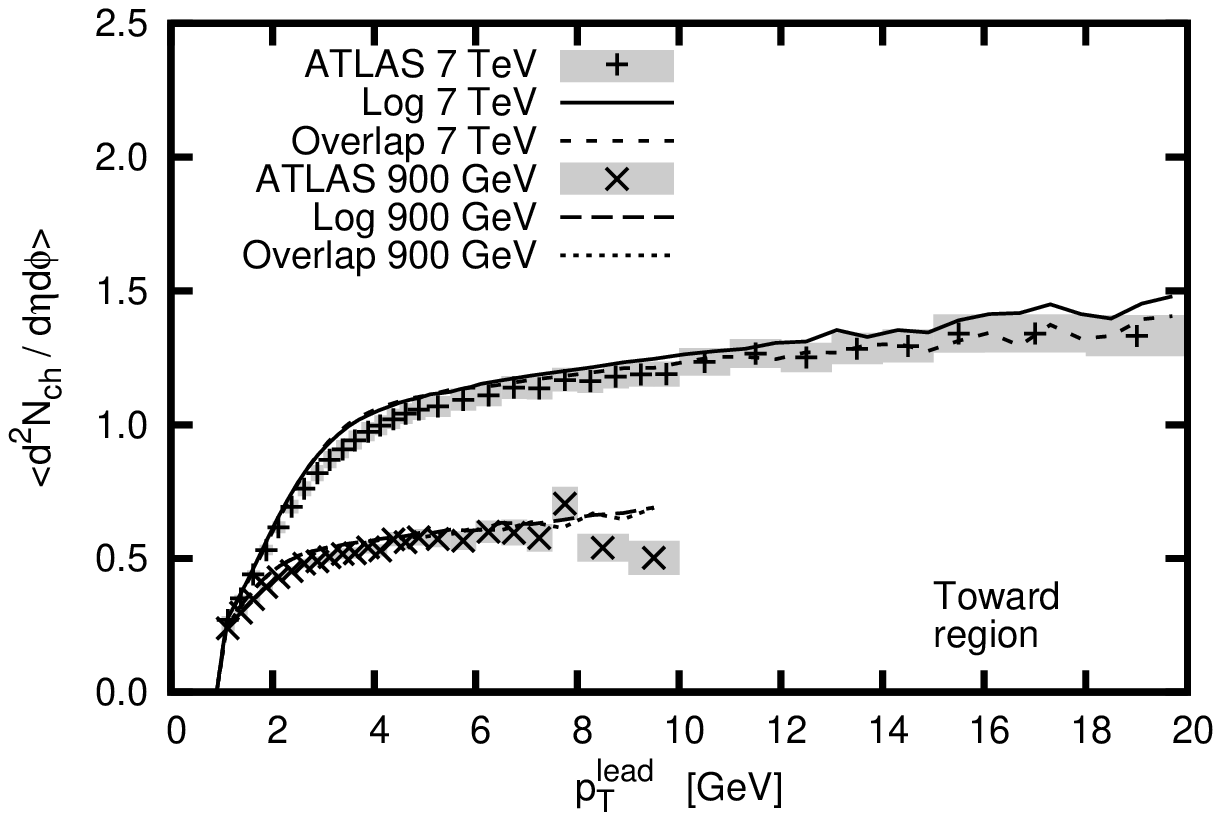}
\includegraphics[scale=0.63,angle=270]{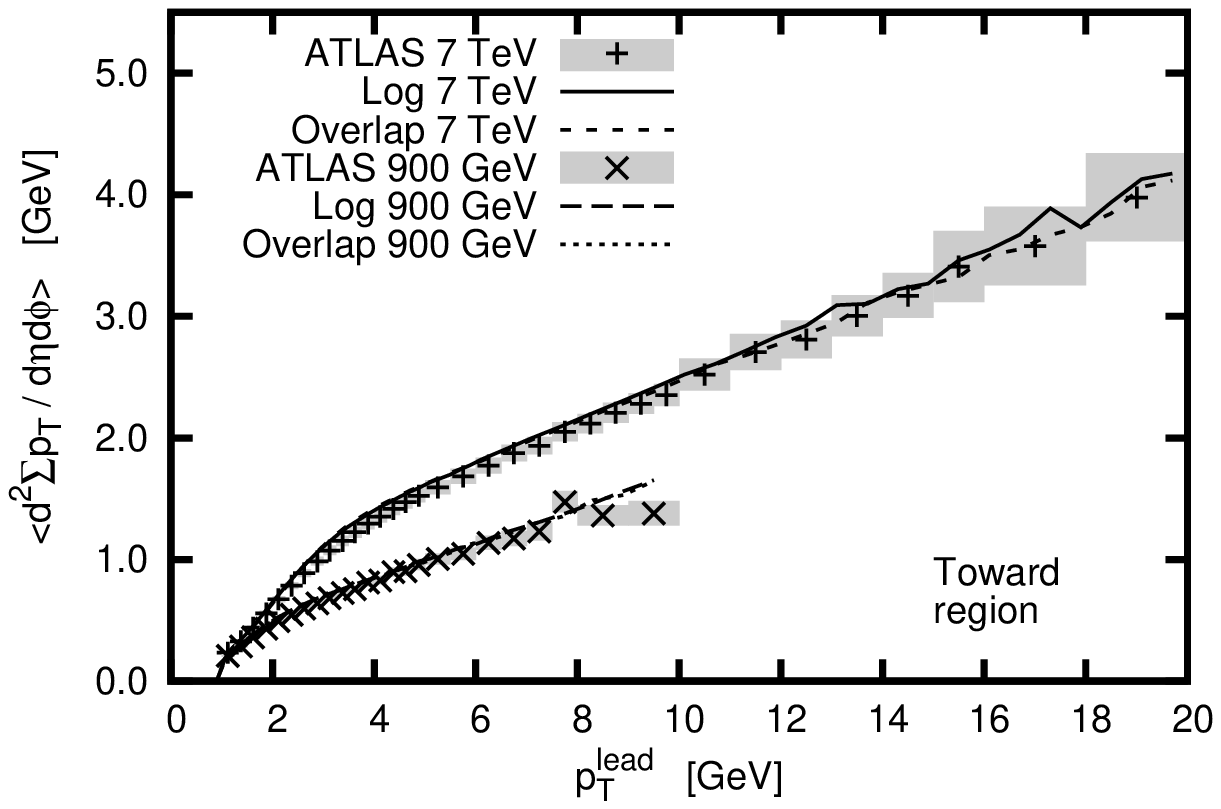}
\includegraphics[scale=0.63,angle=270]{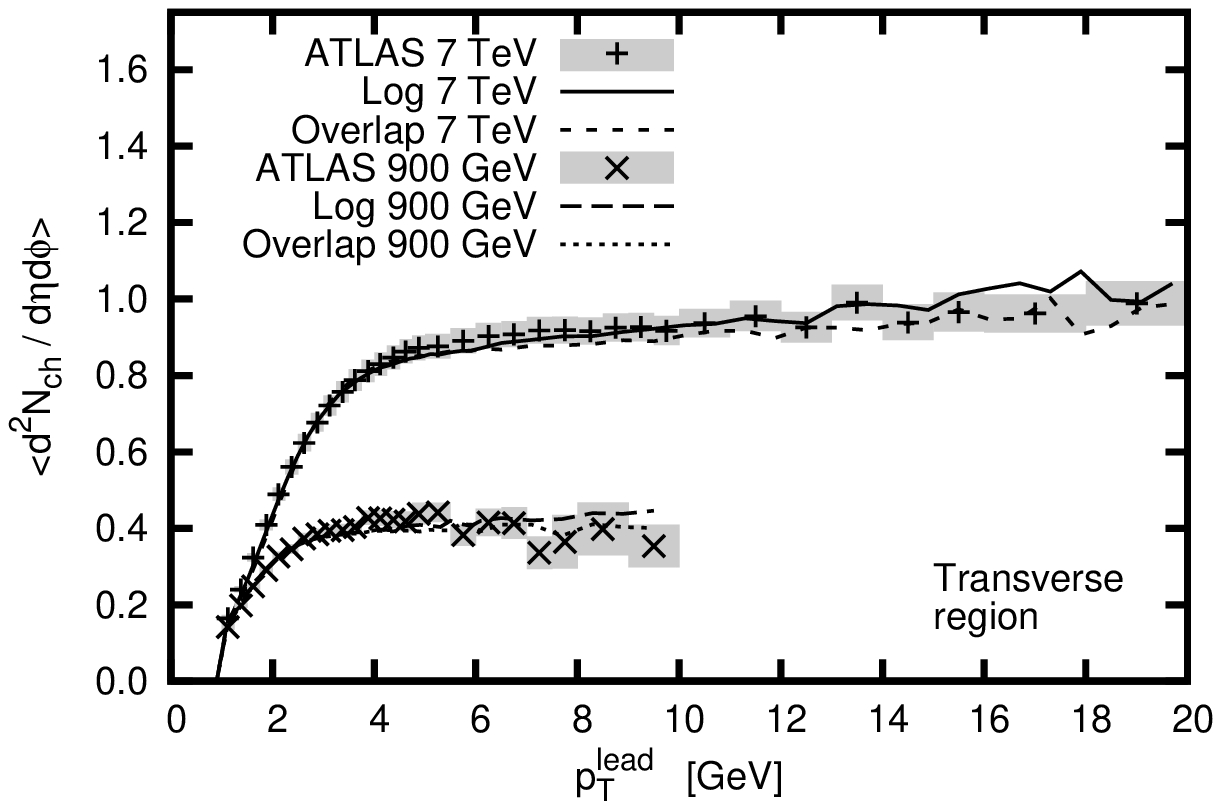}
\includegraphics[scale=0.63,angle=270]{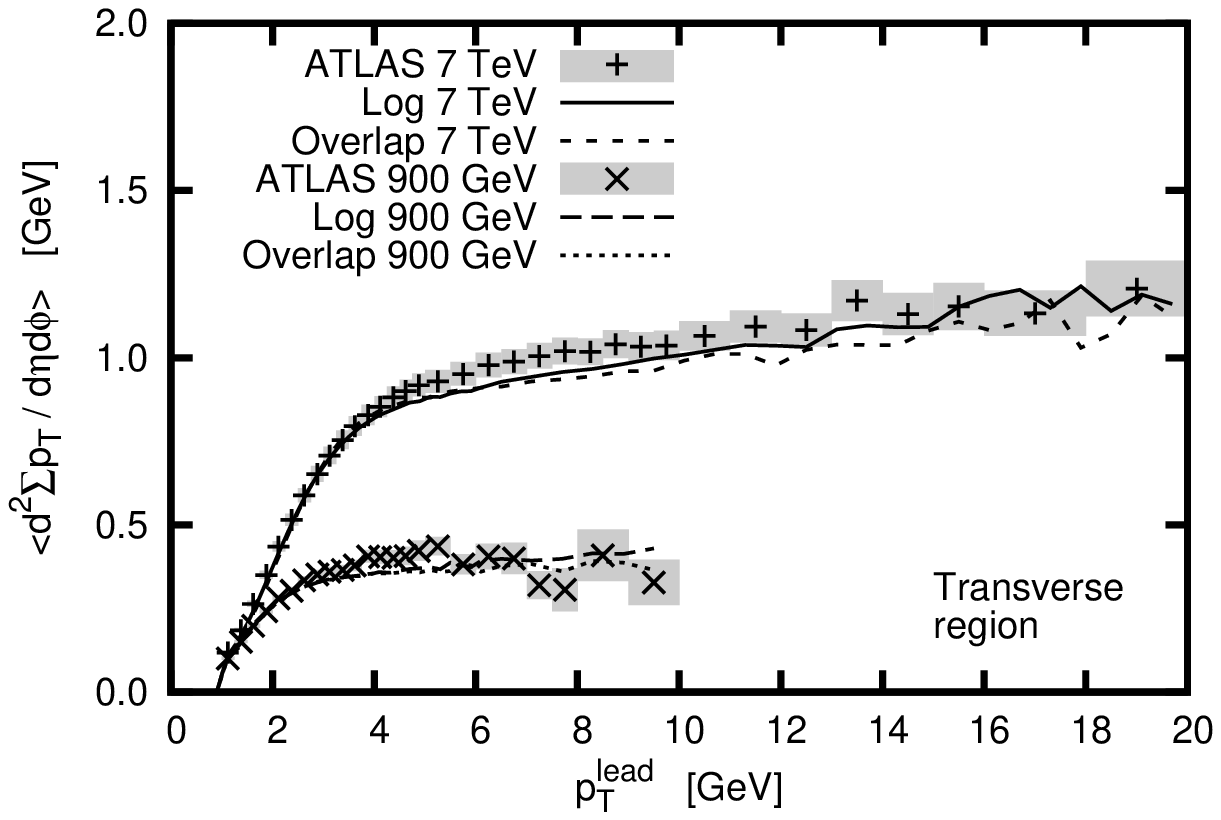}
\caption{Tune 4C, using the log profile, and with a raised $\pTo$ in the
MPI framework, compared against an overlap profile with $p = 1.6$, also
with a raised $\pTo$, and LHC data
\label{fig:4Cmod}}
\end{figure}

The first step is to replace the single Gaussian with the logarithmically
$x$-dependent matter profile with $a_1 = 0.15$. This change is made just in
the non-diffractive component, while that used in diffractive framework for
now remains a single Gaussian. Diffraction primarily impacts the low bins
of the charged multiplicity distribution, and will not greatly affect the
results shown here. Just this change leads to a rise in the tail of the
charged multiplicity distributions, with an increase in activity in all
regions of the underlying event, as expected from the considerations of the
previous sections. This behaviour is most closely matched by an overlap
function with $p = 1.6$, against which we can compare the results. The
simplest way to remove this excess activity is a retuning of the $\pTo$
parameter of the MPI framework, in this case achieved by raising
$\pTo^{\mrm{ref}} = 2.085 \to 2.15 \GeV$.  This rise does not greatly
affect the relative slope of $a_0$, as constrained in Sec.~\ref{sec:psize}.
The results are shown in Fig.~\ref{fig:4Cmod} for the same distributions as
Fig.~\ref{fig:4C}.

After this retuning, the log profile shows some promise. For the charged
multiplicity distribution, the tail now sits above the data with the
overlap profile, while the match is improved with the log.
This effect has already been seen in Fig.~\ref{fig:mbMI}c, where the
overlap and double Gaussian profiles ``shoot up'' in the tails, while the
log profile gives a more gradual rise. The rise of the underlying event is
almost exactly the same in the two different profiles. In the toward
region, the rise is still too fast, and slightly worse than the unmodified
tune. The log profile, here, does have slightly higher tails, consistent
with a narrower matter profile in higher bins of $\pT^{\mrm{lead}}$,
that do suggest a slightly better shape overall. The description in the
transverse region is, in fact, improved, although a further decrease in
activity, for example to improve agreement in the toward region, would
likely push activity lower here, similar to the unmodified tune.

\subsection{Underlying event in Drell-Yan}

Studies can also be made on the underlying event in Drell-Yan processes.
Here, a CDF study \cite{CDFNOTE9351} is used, where a leading $\Zz$ is
reconstructed from the lepton pair. For simplicity, we stay with Tune 4C
and the modified version of it, using the new matter profile, but noting
that it has been shown to give too much activity at the Tevatron. Tune 2C,
also introduced in \cite{Corke:2010yf}, is a tune based only on Tevatron
data. It describes minimum-bias and jet-based underlying event studies
well, but gives only limited agreement with the Drell-Yan results. In
particular, the charged particle number and sum-$\pT$ densities in all
regions are too low. 

\begin{figure}
\centering
\includegraphics[scale=0.6]{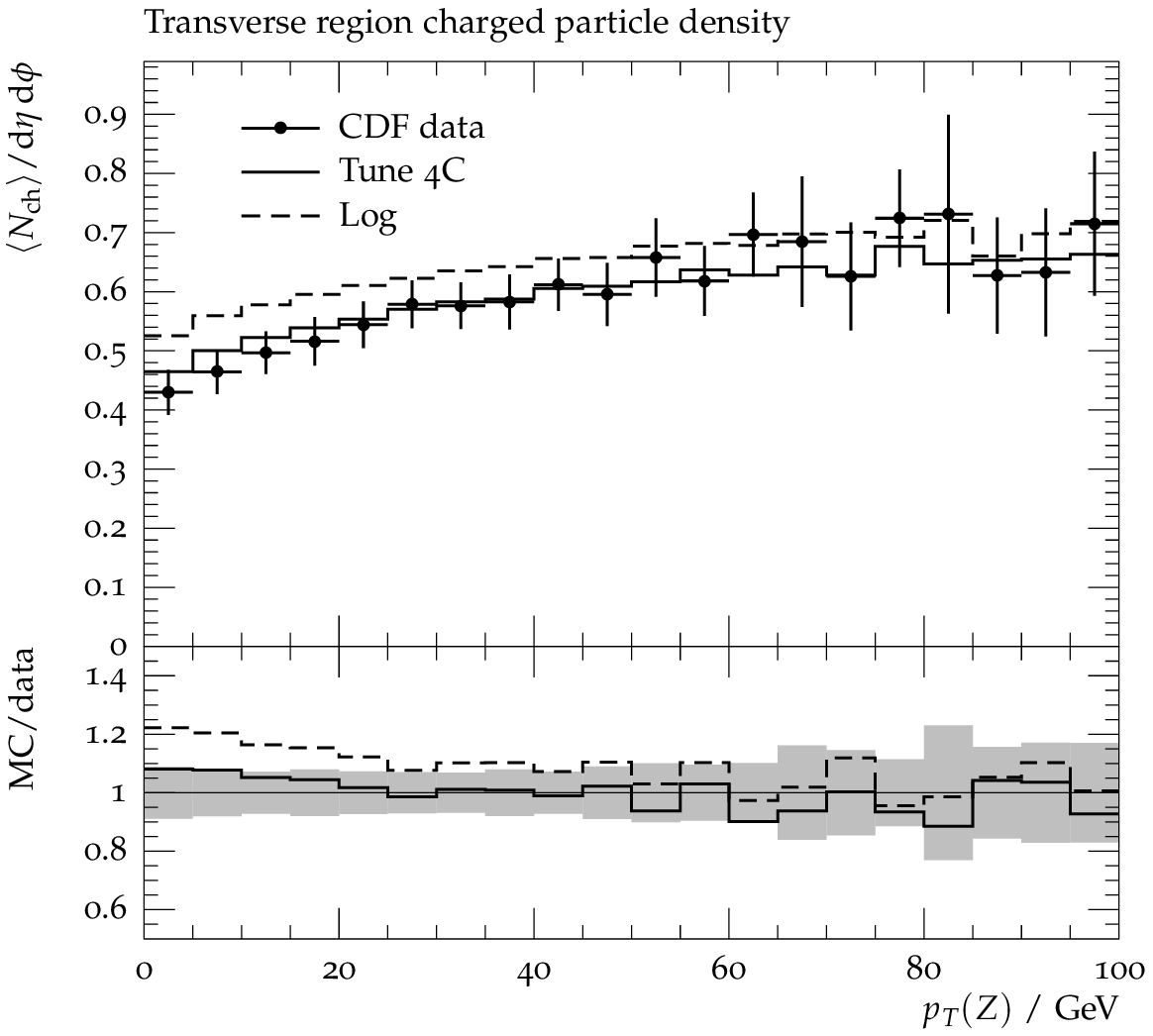}\hspace{4mm}
\includegraphics[scale=0.6]{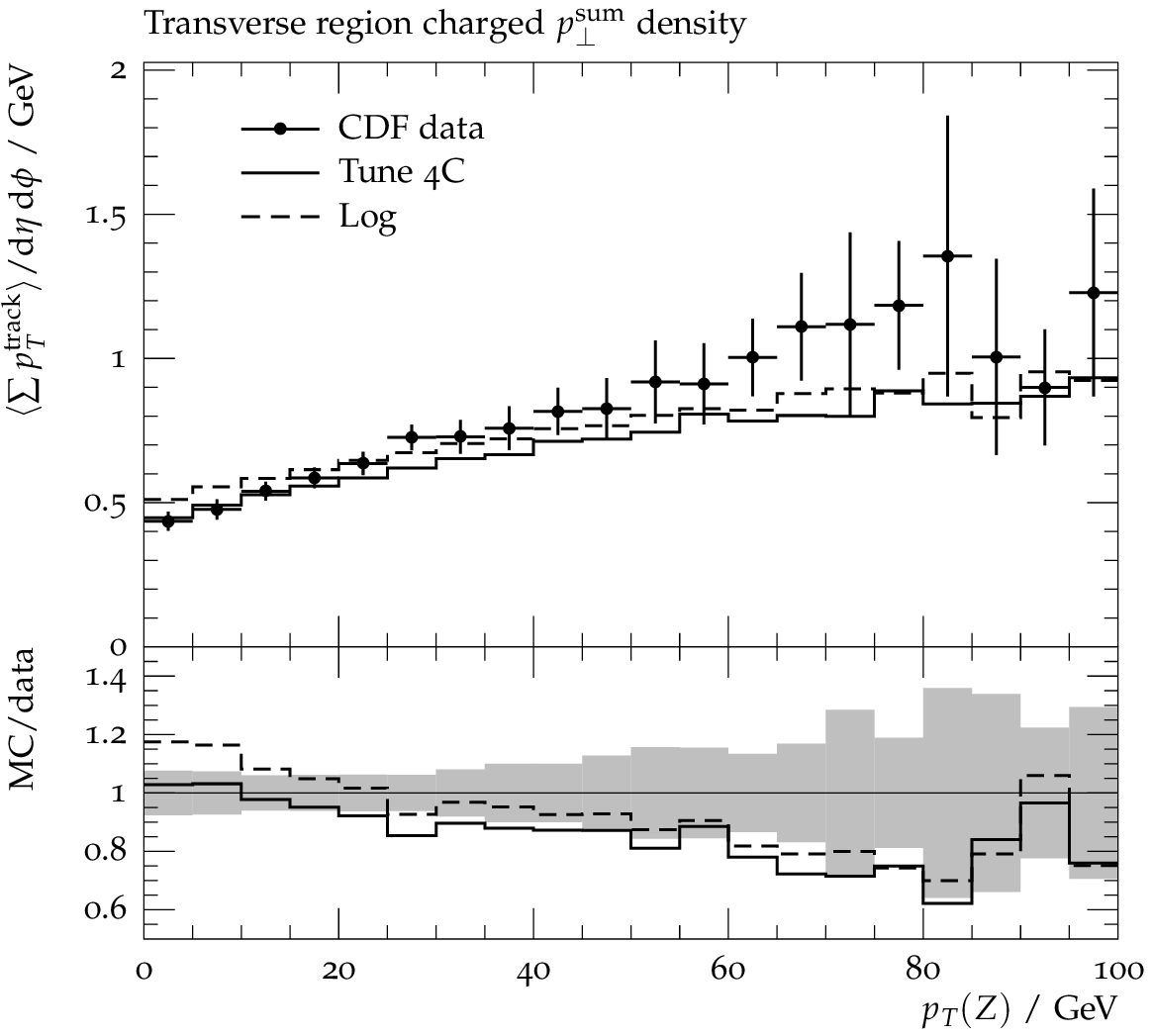}
\caption{Charged number and sum-$\pT$ density in the transverse region of
the underlying event in Drell-Yan
\label{fig:DY}}
\end{figure}

The results from \tsc{Pythia} are run through the Rivet analysis of this
study \cite{Buckley:2010ar}. The charged particle number and sum-$\pT$
densities in the transverse region, as a function of $\pT(\Zz)$, are shown
in Fig.~\ref{fig:DY}. At first glance, it appears that Tune 4C does not do
too badly, but, as above, it is known to give too much activity at the
Tevatron. As expected from previous sections, the log profile, then, gives
an increase in activity, which also occurs in the toward and away regions.
It should be noted that the increase of the sum-$\pT$ density, relative to
the charged number density, is dependent on other factors, such as colour
reconnection, which differ between Tunes 4C and 2C. Given this, the
conclusions we can draw from this study are limited, but the $x$-dependent
matter profile appears to be a step in the right direction.

\section{Summary and outlook}
\label{sec:summary}

There is both theoretical and experimental evidence suggesting that the
wave function of high-$x$ partons should be narrower than that of low-$x$
ones. In this article, we have not tried to examine the underlying
mechanism for this, but instead have modelled the effect using a simple
Gaussian shape with a width that varies logarithmically with the $x$-value
of the parton being probed. This is introduced as a new matter
profile in the MPI framework of \tsc{Pythia 8}.

The framework, outlined in Sec.~\ref{sec:mpi}, is additionally formulated
in terms of a physical size of the proton. Although introduced with a free
parameter, $a_1$, regulating the importance of the
logarithmic component, it can be fixed if it is assumed that the variation
should account for the growth of the total cross section. For this to be
the case, this parameter should lie in the region of $a_1 = 0.15$. In
the studies made here, it has been considered a fixed quantity rather than
a free parameter. The estimates of the proton size come out somewhat below
current low-energy measurements, but as noted, the eikonalisation procedure
neglects the diffractive components of the cross section.

The model gives a matter profile which in some ways is similar to the
double Gaussian scenario. There is an increase in the matter at both small
and large impact parameters, arising naturally due to the form of the
eq.~(\ref{eq:loggauss}). The results are further changed, however, by the
varying enhancement factors in the subsequent chain of MPI.
An early tune to LHC data has been used to quantify the effects of
this profile on minimum-bias and underlying-event distributions. In
particular, although the physical results are somewhat similar to an
intermediate overlap profile, there are differences which give some
indication that this profile could give rise to a viable tune to LHC data.

The case of MPI activity accompanying hard processes which do not contain
final-state particles which can be created in MPI, such that the evolution
covers the entire phase space, is interesting both in its own right, and
as an illustration of the features of the model. The previous matter
profiles give exactly the same impact parameter distribution, regardless of
whether the underlying process is a $\Zz$ or a $1\TeV$ resonance. The new
profile changes this situation. The $x$ dependence now leads to a situation
where the higher-mass resonance will give rise to a narrower impact
parameter profile, leading to changes in both the number of MPI and their
$\pT$ spectrum. In comparisons to data, it leads to extra activity in the
underlying event description of Drell-Yan processes, which appears to be a
step in the right direction, in terms of describing Tevatron data.

The results, then, are promising. A more general tuning to data would help
ascertain more clearly if this profile can improve the overall description,
relative to the other profiles available. Future LHC studies on the
underlying event in Drell-Yan processes would be a welcome addition, in
order to further test the model. The framework will be released in the
upcoming \tsc{Pythia 8.150} version, along with the modified Tune 4C, where
we hope it will be studied further by a wider community.

\subsection*{Acknowledgments}
This work was supported by the Marie Curie Early Stage Training program
``HEP-EST'' (contract number MEST-CT-2005-019626), the Marie
Curie research training network ``MCnet'' (contract number
MRTN-CT-2006-035606), and the Swedish Research Council (contract number
621-2010-3326).

\bibliography{xdep}{}
\bibliographystyle{utcaps}

\end{document}